\documentclass[12pt,a4paper]{report}
\usepackage[brazil]{babel}
\usepackage[latin1]{inputenc}
\usepackage{axodraw}
\usepackage{latexsym}
\usepackage{indentfirst}
\usepackage{graphicx}
\renewcommand{\baselinestretch}{1.2}
\begin{document}
\def\baselinestretch{1.2}
\hoffset=-1.0 true cm
\voffset=-2 true cm
\topmargin=1.0cm
\thispagestyle{empty}
\thicklines
\begin{picture}(370,60)(0,0)
\setlength{\unitlength}{1pt}
\put(40,53){\line(2,3){15}}
\put(40,53){\line(5,6){19}}
\put(40,53){\line(1,1){27}}
\put(40,53){\line(6,5){33}}
\put(40,53){\line(3,2){25}}
\put(40,53){\line(2,1){19}}
\put(40,53){\line(5,-6){17}}
\put(40,53){\line(1,-1){22}}
\put(40,53){\line(6,-5){30}}
\put(40,53){\line(3,-2){22}}
\put(40,53){\line(-2,1){15}}
\put(40,53){\line(-3,1){23}}
\put(40,53){\line(-4,1){26}}
\put(40,53){\line(-6,1){36}}
\put(40,53){\line(-1,0){40}}
\put(40,53){\line(-6,-1){32}}
\put(40,53){\line(-3,-1){20}}
\put(40,53){\line(-2,-1){10}}
\put(75,45){\Huge \bf IFT}
\put(180,56){\small \bf Instituto de F\'\i sica Te\'orica}
\put(165,42){\small \bf Universidade Estadual Paulista} 
\put(-25,2){\line(1,0){433}}
\put(-25,-2){\line(1,0){433}}
\end{picture}

\vskip .3cm
\noindent
{TESE DE DOUTORAMENTO}
\hfill IFT--T.003/06

\vspace{3cm}
\begin{center}
{\large \bf Teorias Semiclássica e Efetiva da Gravitação}

\vspace{1.2cm}
Ricardo Paszko
\end{center}

\vskip 3cm
\hfill Orientador
\vskip 0.4cm
\hfill {\em Prof. Dr. Antonio Accioly}
\vfill
\begin{center}
Novembro de 2006
\end{center}

\setlength{\parindent}{1cm}
\newpage
\pagenumbering{roman}
\begin{center}
{\Large \bf Agradecimentos}
\end{center}
\vskip 2.0cm

Agradeço ao Prof. Accioly por ter acreditado em mim desde o início.

Agradeço à Marcia, Emilia, todos os meus amigos e à minha família pela enorme paciência que tiveram comigo nestes anos de trabalho.

Também agradeço à toda comunidade do IFT pela hospitalidade e à CAPES pelo apoio financeiro integral.

\newpage
\begin{center}
{\Large \bf Resumo}
\end{center}
\vskip 2.0cm

Analisamos o espalhamento de partículas quânticas por um campo gravitacional fraco, tratado como campo externo, em primeira e segunda ordens de perturbação. Essa análise acusa violações do Princípio da Equivalência em relação ao spin --- em primeira ordem ---, e em relação à energia --- em segunda ordem. Verificamos que os resultados mencionados são equivalentes àqueles obtidos por intermédio da Teoria Efetiva da Gravitação, no limite em que uma das massas é muito maior do que as outras energias envolvidas. Discutimos também algumas aplicações de nossa investigação, tais como a determinação de um limite superior para a massa do fóton e a possível detecção em um futuro não muito distante dessas violações do Princípio da Equivalência.
\vskip 1.0cm
\noindent
{\bf Palavras Chaves}: gravitação; teoria semiclássica; teoria efetiva; espalhamento.
\vskip 0.5cm
\noindent
{\bf \'Areas do conhecimento}: Teoria de Campos (1050300-5); Gravitação (1050103-7).

\newpage
\begin{center}
{\Large \bf Abstract}
\end{center}
\vskip 2.0cm

First and second order corrections for the scattering of different types of particles by a weak gravitational field, treated as an external field, are calculated. These computations indicate a violation of the Equivalence Principle: to first order, the cross-sections are spin dependent; if the calculations are pushed to the next order, they become dependent upon energy as well. Interesting enough, the aforementioned results are equivalent to those obtained by means of the so-called Effective Theory of Gravitation, in the limit in which one of the masses is much greater than all the other energies involved. We discuss also some applications of our research, such as the determination of an upper bound for the photon mass, and the possible detection, in the foreseeable future, of these violations of the Equivalence Principle.
\vskip 0.5cm

\vfill \eject

\pagenumbering{arabic}
\tableofcontents

\chapter*{Introdução}
\addcontentsline{toc}{chapter}{Introdução}
\indent Poucos anos depois da Teoria Quântica de Campos (TQC) ser inventada, Léon~Rosenfeld \cite{Rosenfeld} fez a primeira tentativa de quantizar a Relatividade Geral.

Ele seguia uma sugestão de Werner~Heisenberg de que talvez as auto-energias pudessem ser finitas na ausência de campos de matéria. Mas, ao tentar calcular o campo gravitacional produzido por um campo eletromagnético e a auto-energia do gráviton, acabou encontrando os infinitos usuais da TQC.

A primeira tentativa real de quantização foi feita cerca de 20 anos depois por Suraj~Gupta \cite{Gupta52A}, quantizando o campo gravitacional linearizado por meio do mesmo método desenvolvido por ele para a QED (Eletrodinâmica Quântica) \cite{Gupta50}.

Posteriormente, Gupta aplicou o tratamento para o campo não linearizado \cite{Gupta52B}, encontrando divergências mais severas que na QED, e isso sugeria que a teoria não fosse renormalizável.

No início dos anos 60, Richard Feynman descreveu a necessidade dos `ghosts' em cálculos envolvendo loops de grávitons \cite{Feynman63}. Seu argumento foi refinado depois, o que pode ser encontrado nas referências \cite{DeWitt,Faddeev,Feynman72}.

Na mesma época, Christian M\o ller propõe a Teoria Semiclássica da Gravitação, em que a métrica é considerada um campo clássico enquanto todos os outros campos são quantizados \cite{Moller}.

Em 1973, Gerard~'t~Hooft e Martinus~Veltman \cite{'t Hooft} fizeram o primeiro cálculo mostrando que a teoria perturbativa da gravitação não é renormalizável em 1 loop na presença de matéria, mas que poderia ser renormalizável como campo livre.

Mas, em 1985, Marc~Goroff e Augusto~Sagnotti mostraram que a teoria livre não é renormalizável em 2 loops \cite{Goroff}.

Isso fez com que os cálculos perturbativos fossem abandonados, ainda mais com o renascimento da teoria canônica não-perturbativa nos trabalhos de Abhay~Ashtekar \cite{Ashtekar}.

Somente em 1994, nos trabalhos de John~Donoghue \cite{Donoghue}, houve o ressurgimento da teoria perturbativa da gravitação, agora vista como uma teoria efetiva, ou seja, válida em baixas energias.

Nosso trabalho consiste em demonstrar que, em ambas as teorias Semiclássica e Efetiva da Gravitação, existem pequenos desvios do Princípio da Equivalência. Apesar de pequenos, poderão ser observados em um futuro próximo.

Começamos extraindo as Regras de Feynman no Capítulo 1. Já no Capítulo~2, calculamos o espalhamento de diversas partículas em $1^a$ ordem por um campo gravitacional externo fraco. No Capítulo 3, calculamos em $2^a$ ordem e, no capítulo seguinte, verificamos a igualdade entre a Teoria Efetiva e a Teoria Semiclássica. No Capítulo 5, mostramos algumas aplicações para os resultados obtidos e encerramos com a conclusão.

Tentamos manter uma certa continuidade ao longo de nosso trabalho e, por isso, não numeramos as equações. Para isso, nós repetimos algumas expressões e também nomeamos algumas integrais como $I, J, K, \ldots$ nos Apêndices A e B. Os dois últimos apêndices tratam da deflexão gravitacional clássica e semiclássica.

Em todo nosso trabalho usamos as seguintes convenções: $\hbar=c=1$, $\eta_{\mu\nu}=diag(1,-1,-1,-1)$, $\kappa=\sqrt{32\pi G}$ onde $G$ é a constante de Newton e $R=g^{\mu\nu}(\partial_\nu\Gamma^\lambda_{\mu\lambda}-\partial_\lambda\Gamma^\lambda_{\mu\nu}+\Gamma^\tau_{\mu\lambda}\Gamma^\lambda_{\tau\nu}-\Gamma^\tau_{\mu\nu}\Gamma^\lambda_{\tau\lambda})$.

\chapter{Regras de Feynman}
Vamos começar extraindo as Regras de Feynman da Teoria Perturbativa da Gravitação Quântica.

Para isso, escolhemos a seguinte expansão usual da métrica $g_{\mu\nu}$
\[g_{\mu\nu}=\eta_{\mu\nu}+\kappa h_{\mu\nu}\]
onde $\kappa=\sqrt{32\pi G}$, $\eta_{\mu\nu}=diag(1,-1,-1,-1)$ e $h_{\mu\nu}$ representa o campo do gráviton.

A partir desta escolha temos
\begin{eqnarray*}
g^{\mu\nu}&=&\eta^{\mu\nu}-\kappa h^{\mu\nu}+\kappa^2h^{\mu\lambda}h^\nu_\lambda+\ldots\\
\sqrt{-g}&=&1+\frac{\kappa}{2}h+\frac{\kappa^2}{8}(h^2-2h^{\mu\nu}h_{\mu\nu})+\ldots
\end{eqnarray*}
onde definimos $h^{\mu\nu}=\eta^{\mu\alpha}h_{\alpha\beta}\eta^{\beta\nu}$, $h=\eta^{\mu\nu}h_{\mu\nu}$ e $g=\det{g_{\mu\nu}}$.

Com essas definições, podemos atacar o primeiro e mais simples caso: {\it o campo escalar (ou Klein-Gordon) com acoplamento mínimo} \cite{Weinberg}, cuja (densidade de) Lagrangeana é dada por
\[\mathcal{L}_{KG}=\frac{\sqrt{-g}}{2}(g^{\mu\nu}\nabla_\mu\varphi\nabla_\nu\varphi-m^2\varphi^2)\]
onde $\nabla_\mu$ é a {\it derivada covariante} que nesse caso simples se reduz à $\nabla_\mu\varphi=\partial_\mu\varphi$ porque $\varphi$ é um campo escalar.
\newpage
Essa Lagrangeana pode ser expandida em
\begin{eqnarray*}
\mathcal{L}_{KG}&=&\underbrace{\frac{1}{2}(\eta^{\mu\nu}\partial_\mu\varphi\partial_\nu\varphi-m^2\varphi^2)}_{\mathcal{L}_0}\underbrace{-\frac{\kappa}{2}h^{\mu\nu}\left(\partial_\mu\varphi\partial_\nu\varphi-\eta_{\mu\nu}\mathcal{L}_0\right)}_{\mathcal{L}^{(1)}_{Int}}\\&&+\underbrace{\frac{\kappa^2}{2}\left[\frac{1}{4}(h^2-2h^{\mu\nu}h_{\mu\nu})\mathcal{L}_0+\left(h^\mu_\lambda h^{\lambda\nu}-\frac{h}{2}h^{\mu\nu}\right)\partial_\mu\varphi\partial_\nu\varphi\right]}_{\mathcal{L}^{(2)}_{Int}}+\ldots\\
\end{eqnarray*}
onde $\mathcal{L}_0$ é a Lagrangeana livre, ou seja, na ausência de gravidade, $\mathcal{L}^{(1)}_{Int}$ é a Lagrangeana de interação em ordem $\kappa^1$, $\mathcal{L}^{(2)}_{Int}$ é a Lagrangeana de interação em ordem $\kappa^2$ e assim por diante.

A equação de Euler-Lagrange para a Lagrangeana livre
\[\frac{\partial\mathcal{L}_0}{\partial\varphi}-\partial_\mu\frac{\partial\mathcal{L}_0}{\partial(\partial_\mu\varphi)}=0\]
nos dá
\[(\Box+m^2)\varphi(x)=0\]
onde $\Box=\eta^{\mu\nu}\partial_\mu\partial_\nu$. A solução desta equação pode ser escrita como combinações lineares de ondas planas
\[\varphi(x)=\sum_{\bf k}\frac{1}{\sqrt{2VE_k}}\left[a({\bf k})e^{-ik\cdot x}+a^*({\bf k})e^{ik\cdot x}\right]\]
(estamos considerando $\varphi(x)$ como um campo real, portanto $\varphi(x)^*=\varphi(x)$), onde escolhemos uma normalização em volume finito $V$ e $k^\mu=(E_k,{\bf k})$ onde $E_k=\sqrt{{\bf k}^2+m^2}$.

Quantizando o campo $\varphi(x)$ através das relações de comutação em tempos iguais (veja por exemplo \cite{Mandl})
\begin{eqnarray*}
&[\varphi({\bf x},t),\pi({\bf x'},t)]=i\delta^{(3)}({\bf x-x'})&\\
&[\varphi({\bf x},t),\varphi({\bf x'},t)]=[\pi({\bf x},t),\pi({\bf x'},t)]=0&
\end{eqnarray*}
onde
\[\pi(x)\equiv\frac{\partial\mathcal{L}_0}{\partial\dot\varphi}=\dot\varphi(x)\]
promovemos $a({\bf k})$ e $a^*({\bf k})$ aos operadores
\begin{eqnarray*}
\begin{array}{cccc}
a({\bf k})\vert 0\rangle&=&0&\mbox{aniquilação}\\
a^\dagger({\bf k})\vert 0\rangle&=&\vert{\bf k}\rangle&\mbox{criação}
\end{array}
\end{eqnarray*}
onde $\vert 0\rangle$ é o estado de vácuo, e assim obtemos
\begin{eqnarray*}
&[a({\bf k}),a^\dagger({\bf k'})]=i\delta_{kk'}&\\
&[a({\bf k}),a({\bf k'})]=[a^\dagger({\bf k}),a^\dagger({\bf k'})]=0.&
\end{eqnarray*}

Com isso, podemos calcular o propagador do campo escalar (veja detalhes em \cite{Mandl})
\[\langle 0\vert T\{\varphi (x)\varphi (x')\}\vert 0\rangle=\int\frac{d^4p}{(2\pi)^4}e^{-ip\cdot (x-x')}\frac{i}{p^2-m^2+i\epsilon}\]
onde $T$ é a ordem temporal.

\noindent{\bf Propagador do escalar}
\begin{center}
\begin{picture}(100,20)(0,0)
\DashArrowLine(0,10)(60,10){5}
\Text(30,5)[ct]{$p$}
\Text(80,10)[l]{\Large $\frac{i}{p^2-m^2+i\epsilon}$}
\end{picture}
\end{center}

De modo análogo, o campo real $h_{\mu\nu}(x)$ também pode ser quantizado
\[h^{\mu\nu}(x)=\sum_{{\bf k},r}\frac{\epsilon^{\mu\nu}_r({\bf k})}{\sqrt{2VE_k}}\left[b_r({\bf k})e^{-ik\cdot x}+b_r^\dagger({\bf k})e^{ik\cdot x}\right]\]
onde $\epsilon^{\mu\nu}_r({\bf k})$ são os tensores de polarização (similares aos do caso do fóton, veja mais detalhes em \cite{Gupta52A,Weinberg64}).

Tendo em mãos as expressões dos campos quantizados, podemos agora extrair o primeiro vértice a partir de $\mathcal{L}^{(1)}_{Int}$ através da matriz $S$ dada por
\[S=1\underbrace{-i\int d^4x:\mathcal{H}^{(1)}_{Int}(x):}_{S^{(1)}}+\ldots\]
onde $\mathcal{H}^{(1)}_{Int}(x)=-\mathcal{L}^{(1)}_{Int}(x)$ é a Hamiltoniana de interação (em $1^a$ ordem) e $:O:$ é o ordenamento normal do operador $O$.

Considere o estado inicial formado por um escalar com momento ${\bf p}$ e um gráviton com momento ${\bf q}$ e polarização $r$
\[\vert i\rangle=a^\dagger({\bf p})b_r^\dagger({\bf q})\vert 0\rangle\]
já o estado final apenas um escalar com momento ${\bf p'}$
\[\vert f\rangle=a^\dagger({\bf p'})\vert 0\rangle\]
portanto
\begin{eqnarray*}
\langle f\vert S^{(1)}\vert i\rangle&=&\langle 0\vert a({\bf p'})S^{(1)}a^\dagger({\bf p})b_r^\dagger({\bf q})\vert 0\rangle\\
&=&\langle 0\vert a({\bf p'})i\int d^4x:\mathcal{L}^{(1)}_{Int}(x):a^\dagger({\bf p})b_r^\dagger({\bf q})\vert 0\rangle\\
&=&-\frac{i\kappa}{2}\langle 0\vert a({\bf p'})\int d^4x:h^{\mu\nu}\left(\partial_\mu\varphi\partial_\nu\varphi-\eta_{\mu\nu}\mathcal{L}_0\right)
:a^\dagger({\bf p})b_r^\dagger({\bf q})\vert 0\rangle\\
\end{eqnarray*}
que após uma simples álgebra de operadores resulta em
\begin{eqnarray*}
\langle f\vert S^{(1)}\vert i\rangle&=&\frac{(2\pi)^4\delta^{(4)}(p'-p-q)\epsilon^{\mu\nu}_r({\bf q})}{\sqrt{2VE_{p'}}\sqrt{2VE_p}\sqrt{2VE_q}}\times\\
&&\underbrace{-\frac{i\kappa}{2}[p'_\mu p_\nu+p'_\nu p_\mu-\eta_{\mu\nu}(p'\cdot p-m^2)]}_{\equiv V_{\mu\nu}(p',p)}
\end{eqnarray*}
e ignorando fatores de normalização, bem como o tensor de polarização e o delta de Dirac, obtemos nosso primeiro vértice:

\noindent{\bf Vértice gráviton - escalar - escalar}
\begin{center}
\begin{picture}(90,104)(0,0)
\Gluon(0,52)(60,52){3}{6}
\LongArrow(25,59)(35,59)
\Text(60,42)[r]{$\alpha\beta$}
\Text(30,42)[c]{$q$}
\DashArrowLine(60,52)(90,104){5}
\Text(80,78)[l]{$p'=p+q$}
\DashArrowLine(90,0)(60,52){5}
\Text(80,26)[l]{$p$}
\end{picture}
\end{center}
\begin{eqnarray*}
V^{\alpha\beta}(p',p)=-\frac{i\kappa}{2}[p'^\alpha p^\beta+p'^\beta p^\alpha-\eta^{\alpha\beta}(p'\cdot p-m^2)]
\end{eqnarray*}
e de maneira análoga obtemos os vértices de ordem mais alta:

\noindent{\bf Vértice gráviton - gráviton - escalar - escalar}
\begin{center}
\begin{picture}(84,84)(0,0)
\Gluon(0,0)(42,42){3}{6}
\LongArrow(30.5,17.5)(38.5,24.5)
\Text(9,21)[r]{$k$}
\Text(14,2)[l]{$\alpha\beta$}
\DashArrowLine(84,0)(42,42){5}
\Text(75,21)[l]{$p$}
\Gluon(42,42)(0,84){-3}{6}
\LongArrow(38.5,59.5)(30.5,66.5)
\Text(9,63)[r]{$k-q$}
\Text(14,81)[l]{$\mu\nu$}
\DashArrowLine(42,42)(84,84){5}
\Text(75,63)[l]{$p'=p+q$}
\end{picture}
\end{center}
\begin{eqnarray*}
V^{\mu\nu,\alpha\beta}(p',p)&=&i\kappa^2\left\{\left[I^{\mu\nu,\lambda\sigma}I_\sigma^{\rho,\alpha\beta}-\frac{1}{4}(\eta^{\mu\nu}I^{\alpha\beta,\lambda\rho}+\eta^{\alpha\beta}I^{\mu\nu,\lambda\rho})\right](p'_\lambda p_\rho+p'_\rho p_\lambda)\right.\\&&\left.-\frac{1}{2}(p'\cdot p-m^2)\mathcal{P}^{\alpha\beta,\mu\nu}\right\}
\end{eqnarray*}
onde $\mathcal{P}_{\alpha\beta,\mu\nu}=I_{\alpha\beta,\mu\nu}-\frac{1}{2}\eta_{\alpha\beta}\eta_{\mu\nu}$ e $I_{\alpha\beta,\mu\nu}=\frac{1}{2}(\eta_{\alpha\mu}\eta_{\beta\nu}+\eta_{\alpha\nu}\eta_{\beta\mu})$.

Para o campo de Dirac temos
\begin{eqnarray*}
\mathcal{L}_D&=&\sqrt{-g}\left[\frac{i}{2}(\bar\psi\gamma^\mu\overrightarrow\nabla_\mu\psi-\bar\psi\overleftarrow\nabla_\mu\gamma^\mu\psi)-m\bar\psi\psi\right]\\&=&\underbrace{\frac{i}{2}(\bar\psi\overrightarrow{\not\partial}\psi-\bar\psi\overleftarrow{\not\partial} \psi)-m\bar\psi\psi}_{\mathcal{L}_0}-\frac{\kappa}{2}h^{\mu\nu}\left[\frac{i}{2}(\bar\psi\gamma_\mu\partial_\nu\psi-\partial_\nu\bar\psi\gamma_\mu\psi)-\eta_{\mu\nu}\mathcal{L}_0\right]+\ldots
\end{eqnarray*}
onde $\overrightarrow\nabla_\mu\psi=\partial_\mu\psi+iw_\mu\psi$, $\bar\psi\overleftarrow\nabla_\mu=\partial_\mu\bar\psi-i\bar\psi w_\mu$ e a conexão $w_\mu$ pode ser escrita em função da tetrada $e^\mu_m$ (e sua inversa $e_\mu^m$) como \cite{Choi,Weinberg}
\[w_\mu=\frac{1}{4}\sigma^{mn}\left[e^\nu_m(\partial_\mu e_{n\nu}-\partial_\nu e_{n\mu})+\frac{1}{2}e^\rho_me^\sigma_n(\partial_\sigma e_{l\rho}-\partial_\rho e_{l\sigma})e^l_\mu-(m\leftrightarrow n)\right]\]
e a tetrada pode ser expandida em $e^m_\mu=\delta^m_\mu+\frac{\kappa}{2}h^m_\mu+\ldots$, portanto

\noindent{\bf Propagador do elétron}
\begin{center}
\begin{picture}(100,20)(0,0)
\ArrowLine(0,10)(60,10)
\Text(30,5)[ct]{$p$}
\Text(80,10)[l]{\Large $\frac{i}{\not p-m+i\epsilon}$}
\end{picture}
\end{center}

\noindent{\bf Vértice gráviton - elétron - elétron}
\begin{center}
\begin{picture}(90,104)(0,0)
\Gluon(0,52)(60,52){3}{6}
\LongArrow(25,59)(35,59)
\Text(60,42)[r]{$\alpha\beta$}
\Text(30,42)[c]{$q$}
\ArrowLine(60,52)(90,104)
\Text(80,78)[l]{$p'=p+q$}
\ArrowLine(90,0)(60,52)
\Text(80,26)[l]{$p$}
\end{picture}
\end{center}
\begin{eqnarray*}
V^{\alpha\beta}(p',p)=\frac{i}{8}\kappa[2\eta^{\alpha\beta}(\not p'+\not p-2m)-(p'+p)^\alpha\gamma^\beta-(p'+p)^\beta\gamma^\alpha].
\end{eqnarray*}
Vértices de ordem mais alta no caso de férmions não serão usados nesta tese.

Para o campo de Maxwell temos
\begin{eqnarray*}
\mathcal{L}_M&=&-\frac{\sqrt{-g}}{4}g^{\alpha\beta}g^{\mu\nu}F_{\alpha\mu}F_{\beta\nu}\\&=&\underbrace{-\frac{1}{4}\eta^{\alpha\beta}\eta^{\mu\nu}F_{\alpha\mu}F_{\beta\nu}}_{\mathcal{L}_0}-\frac{\kappa}{2}h^{\mu\nu}\left[-\eta^{\alpha\beta}F_{\alpha\mu}F_{\beta\nu}-\eta_{\mu\nu}\mathcal{L}_0\right]\\&&+\frac{\kappa^2}{4}\left[\frac{1}{2}(h^2-2h^{\mu\nu}h_{\mu\nu})\mathcal{L}_0+F_{\alpha\beta}F_{\mu\nu}(hh^{\alpha\mu}\eta^{\beta\nu}-2h^{\alpha\lambda}h^\mu_\lambda\eta^{\beta\nu}-h^{\alpha\mu}h^{\beta\nu})\right]+\ldots
\end{eqnarray*}
onde $F_{\mu\nu}=\partial_\mu A_\nu-\partial_\nu A_\mu$, $A_\mu$ é o campo do fóton e escolhendo o gauge de Feynman, $\partial^\mu A_\mu=0$, temos

\noindent{\bf Propagador do fóton}
\begin{center}
\begin{picture}(100,20)(0,0)
\Photon(0,10)(60,10){3}{6}
\LongArrow(25,17)(35,17)
\Text(0,0)[l]{$\alpha$}
\Text(60,0)[r]{$\beta$}
\Text(30,0)[c]{$q$}
\Text(80,10)[l]{\Large $\frac{-i\eta_{\alpha\beta}}{q^2+i\epsilon}$}
\end{picture}
\end{center}

\noindent{\bf Vértice gráviton - fóton - fóton}
\begin{center}
\begin{picture}(120,120)(0,0)
\Gluon(0,60)(60,60){-3}{6}
\LongArrow(25,67)(35,67)
\Text(30,50)[c]{$q$}
\Text(56,50)[r]{$\alpha\beta$}
\Photon(90,111.96)(60,60){3}{6}
\LongArrow(62.5,81.65)(66.5,90.31)
\Text(90,85.98)[l]{$p'=p+q$}
\Text(82,105.96)[rb]{$\nu$}
\Photon(90,8.04)(60,60){3}{6}
\LongArrow(67.5,29.69)(62.5,38.35)
\Text(90,34.02)[l]{$p$}
\Text(82,14.04)[rt]{$\mu$}
\end{picture}
\end{center}
\begin{eqnarray*}
V^{\alpha\beta,\mu\nu}(p',p)&=&-\frac{i\kappa}{2}[(\eta^{\alpha\beta}\eta^{\mu\nu}-\eta^{\alpha\mu}\eta^{\beta\nu}-\eta^{\alpha\nu}\eta^{\beta\mu})p'\cdot p-\eta^{\alpha\beta}p'^\mu p^\nu+\eta^{\mu\beta}p'^\alpha p^\nu\\&&-\eta^{\mu\nu}p'^\alpha p^\beta+\eta^{\alpha\nu}p'^\mu p^\beta+\eta^{\beta\nu}p'^\mu p^\alpha-\eta^{\mu\nu}p'^\beta p^\alpha+\eta^{\alpha\mu}p'^\beta p^\nu]
\end{eqnarray*}

\noindent{\bf Vértice gráviton - gráviton - fóton - fóton}
\begin{center}
\begin{picture}(84,84)(0,0)
\Gluon(0,0)(42,42){3}{6}
\LongArrow(30.5,17.5)(38.5,24.5)
\Text(9,21)[r]{$k$}
\Text(14,2)[l]{$\alpha\beta$}
\Photon(84,0)(42,42){3}{6}
\LongArrow(53.5,17.5)(45.5,24.5)
\Text(75,21)[l]{$p$}
\Text(64,2)[l]{$\lambda$}
\Gluon(42,42)(0,84){-3}{6}
\LongArrow(38.5,59.5)(30.5,66.5)
\Text(9,63)[r]{$k-q$}
\Text(14,81)[l]{$\mu\nu$}
\Photon(42,42)(84,84){3}{6}
\LongArrow(45.5,59.5)(53.5,66.5)
\Text(75,63)[l]{$p'=p+q$}
\Text(64,82)[l]{$\rho$}
\end{picture}
\end{center}
\begin{eqnarray*}
V_{\alpha\beta,\mu\nu,\lambda\rho}(p',p)&=&-\frac{i}{4}\kappa^2[(\eta_{\alpha\beta}\eta_{\mu\nu}-2\eta_{\alpha\mu}\eta_{\beta\nu})(\eta_{\lambda\rho}p \cdot p'-p_\rho p'_\lambda)\\&&-\eta_{\alpha\beta}(T_{\lambda\rho\mu\nu}+T_{\lambda\rho\nu\mu})-\eta_{\mu\nu}(T_{\lambda\rho\alpha\beta}+T_{\lambda\rho\beta\alpha})\\&&+2\eta_{\beta\mu}(T_{\lambda\rho\alpha\nu}+T_{\lambda\rho\nu\alpha})+2\eta_{\alpha\nu}(T_{\lambda\rho\mu\beta}+T_{\lambda\rho\beta\mu})\\&&+2(\eta_{\mu\lambda}\eta_{\rho\nu}p_\alpha p'_\beta-\eta_{\lambda\alpha}\eta_{\rho\nu}p_\mu p'_\beta-\eta_{\lambda\mu}\eta_{\rho\beta}p_\alpha p'_\nu\\&&+\eta_{\lambda\nu}\eta_{\rho\mu}p_\beta p'_\alpha-\eta_{\lambda\beta}\eta_{\rho\mu}p_\nu p'_\alpha-\eta_{\lambda\nu}\eta_{\rho\alpha}p_\beta p'_\mu\\&&+\eta_{\lambda\alpha}\eta_{\rho\beta}p_\mu p'_\nu+\eta_{\lambda\beta}\eta_{\rho\alpha}p_\nu p'_\mu)]
\end{eqnarray*}
onde $T_{\alpha\beta\mu\nu}=\eta_{\alpha\beta}p_\mu p'_\nu-\eta_{\alpha\mu}p_\beta p'_\nu-\eta_{\beta\nu}p_\mu p'_\alpha+\eta_{\alpha\mu}\eta_{\beta\nu}p\cdot p'$.

Analogamente para o campo de Proca
\begin{eqnarray*}
\mathcal{L}_P&=&\sqrt{-g}\left(-\frac{1}{4}g^{\alpha\beta}g^{\mu\nu}F_{\alpha\mu}F_{\beta\nu}+\frac{m^2}{2}g^{\alpha\beta}A_\alpha A_\beta\right)\\&=&\underbrace{-\frac{1}{4}\eta^{\alpha\beta}\eta^{\mu\nu}F_{\alpha\mu}F_{\beta\nu}+\frac{m^2}{2}\eta^{\alpha\beta}A_\alpha A_\beta}_{\mathcal{L}_0}\\&&-\frac{\kappa}{2}h^{\mu\nu}\left[-\eta^{\alpha\beta}F_{\alpha\mu}F_{\beta\nu}+m^2A_\mu A_\nu-\eta_{\mu\nu}\mathcal{L}_0\right]+\ldots
\end{eqnarray*}
e

\noindent{\bf Vértice gráviton - fóton massivo - fóton massivo}
\begin{center}
\begin{picture}(120,120)(0,0)
\Gluon(0,60)(60,60){-3}{6}
\LongArrow(25,67)(35,67)
\Text(30,50)[c]{$q$}
\Text(56,50)[r]{$\alpha\beta$}
\Photon(90,111.96)(60,60){3}{6}
\LongArrow(62.5,81.65)(66.5,90.31)
\Text(90,85.98)[l]{$p'=p+q$}
\Text(82,105.96)[rb]{$\nu$}
\Photon(90,8.04)(60,60){3}{6}
\LongArrow(67.5,29.69)(62.5,38.35)
\Text(90,34.02)[l]{$p$}
\Text(82,14.04)[rt]{$\mu$}
\end{picture}
\end{center}
\begin{eqnarray*}
V^{\alpha\beta,\mu\nu}(p',p)&=&-\frac{i\kappa}{2}[(\eta^{\alpha\beta}\eta^{\mu\nu}-\eta^{\alpha\mu}\eta^{\beta\nu}-\eta^{\alpha\nu}\eta^{\beta\mu})(p'\cdot p-m^2)\\&&-\eta^{\alpha\beta}p'^\mu p^\nu+\eta^{\mu\beta}p'^\alpha p^\nu-\eta^{\mu\nu}p'^\alpha p^\beta+\eta^{\alpha\nu}p'^\mu p^\beta\\&&+\eta^{\beta\nu}p'^\mu p^\alpha-\eta^{\mu\nu}p'^\beta p^\alpha+\eta^{\alpha\mu}p'^\beta p^\nu].
\end{eqnarray*}
Também não vamos precisar de mais vértices com fótons massivos.

Para o campo do gráviton, escolhendo ainda o gauge de de Donder, ou seja, $\partial_\mu h^\mu_\nu=\frac{1}{2}\partial_\nu h$, obtemos a seguinte expansão da Lagrangeana de Einstein-Hilbert
\begin{eqnarray*}
\mathcal{L}_{EH}&=&\frac{2}{\kappa^2}\sqrt{-g}R\\&=&\frac{1}{2}\left(\partial^\mu h^{\nu\lambda}\partial_\mu h_{\nu\lambda}-\frac{1}{2}\partial^\mu h\partial_\mu h\right)+\kappa\left(\frac{1}{2}h^\alpha_\beta\partial^\mu h^\beta_\alpha\partial_\mu h-\frac{1}{2}h^\alpha_\beta\partial_\alpha h^\mu_\nu\partial^\beta h^\nu_\mu\right.\\&&-\left.h^\alpha_\beta\partial_\mu h^\nu_\alpha\partial^\mu h^\beta_\nu+\frac{1}{4}h\partial^\alpha h^\mu_\nu\partial_\alpha h^\nu_\mu+h^\beta_\mu\partial_\nu h^\alpha_\beta\partial^\mu h^\nu_\alpha-\frac{1}{8}h\partial^\mu h\partial_\mu h\right)+\ldots
\end{eqnarray*}
na expressão acima descartamos divergências totais e deixamos o segundo termo o mais compacto possível \cite{Choi}. Com isto podemos retirar o propagador do gráviton e o primeiro vértice:

\noindent{\bf Propagador do gráviton}
\begin{center}
\begin{picture}(100,20)(0,0)
\Gluon(0,10)(60,10){3}{6}
\LongArrow(25,17)(35,17)
\Text(0,0)[l]{$\alpha\beta$}
\Text(60,-1)[r]{$\mu\nu$}
\Text(30,0)[c]{$q$}
\Text(80,10)[l]{\Large $\frac{i\mathcal{P}_{\alpha\beta,\mu\nu}}{q^2+i\epsilon}$}
\end{picture}
\end{center}
onde $\mathcal{P}_{\alpha\beta,\mu\nu}=I_{\alpha\beta,\mu\nu}-\frac{1}{2}\eta_{\alpha\beta}\eta_{\mu\nu}$, $I_{\alpha\beta,\mu\nu}=\frac{1}{2}(\eta_{\alpha\mu}\eta_{\beta\nu}+\eta_{\alpha\nu}\eta_{\beta\mu})$ e

\noindent{\bf Vértice gráviton - gráviton - gráviton}
\begin{center}
\begin{picture}(120,120)(0,0)
\Gluon(60,60)(120,60){-3}{6}
\LongArrow(85,67)(95,67)
\Text(90,50)[c]{$q$}
\Text(120,49)[r]{$\mu\nu$}
\Gluon(30,111.96)(60,60){3}{6}
\LongArrow(59.5,81.65)(55.5,90.31)
\Text(35,85.98)[r]{$k-q$}
\Text(36,105.96)[lb]{$\lambda\rho$}
\Gluon(30,8.04)(60,60){3}{6}
\LongArrow(52.5,29.69)(57.5,38.35)
\Text(35,34.02)[r]{$k$}
\Text(36,14.04)[lt]{$\alpha\beta$}
\end{picture}
\end{center}
\begin{eqnarray*}
V_{\alpha\beta,\mu\nu,\lambda\rho}(k,q,k-q)&=&i\kappa\left\{-\frac{1}{2}(k^2+q^2+(k-q)^2)\left[I^\sigma_{\mu,\alpha\beta}I_{\lambda\rho,\sigma\nu}+I^\sigma_{\nu,\alpha\beta}I_{\lambda\rho,\sigma\mu}\right.\right.\\&&\left.\left.+\frac{1}{4}\eta_{\alpha\beta}\eta_{\mu\nu}\eta_{\lambda\rho}-\frac{1}{2}(\eta_{\alpha\beta}I_{\mu\nu,\lambda\rho}+\eta_{\mu\nu}I_{\alpha\beta,\lambda\rho}+\eta_{\lambda\rho}I_{\alpha\beta,\mu\nu})\right]\right.\\&&\left.+q^\sigma(k-q)^\tau\left[I_{\mu\nu,\lambda\rho}I_{\alpha\beta,\sigma\tau}-\frac{1}{2}(I_{\alpha\beta,\rho\sigma}I_{\mu\nu,\lambda\tau}+I_{\alpha\beta,\lambda\sigma}I_{\mu\nu,\rho\tau}\right.\right.\\&&\left.\left.+I_{\lambda\rho,\nu\sigma}I_{\alpha\beta,\mu\tau}+I_{\lambda\rho,\mu\sigma}I_{\alpha\beta,\nu\tau})\right]\right.\\&&\left.-k^\sigma(k-q)^\tau\left[I_{\alpha\beta,\lambda\rho}I_{\mu\nu,\sigma\tau}-\frac{1}{2}(I_{\mu\nu,\rho\sigma}I_{\alpha\beta,\lambda\tau}+I_{\mu\nu,\lambda\sigma}I_{\alpha\beta,\rho\tau}\right.\right.\\&&\left.\left.+I_{\lambda\rho,\alpha\sigma}I_{\mu\nu,\beta\tau}+I_{\lambda\rho,\beta\sigma}I_{\mu\nu,\alpha\tau})\right]\right.\\&&\left.-k^\sigma q^\tau\left[I_{\alpha\beta,\mu\nu}I_{\lambda\rho,\sigma\tau}-\frac{1}{2}(I_{\lambda\rho,\beta\tau}I_{\mu\nu,\alpha\sigma}+I_{\lambda\rho,\alpha\tau}I_{\mu\nu,\beta\sigma}\right.\right.\\&&\left.\left.+I_{\lambda\rho,\nu\sigma}I_{\alpha\beta,\mu\tau}+I_{\lambda\rho,\mu\sigma}I_{\alpha\beta,\nu\tau})\right]\right\}.
\end{eqnarray*}

\chapter{Campo Externo - $1^a$ ordem}
Considere o espalhamento de uma partícula quantizada de massa $m$ (possivelmente nula), spin $s$ e (tri)momento inicial ${\bf p}$ (e final ${\bf p'}$) por um campo externo clássico dado por
\[h_{\mu\nu}(x)=\frac{2GM}{\kappa |{\bf x}|}(\eta_{\mu\nu}-2\eta_{\mu 0}\eta_{\nu 0})\]
que é a solução da equação de Einstein linearizada no gauge de de Donder
\[\Box\left[h_{\mu\nu}(x)-\frac{1}{2}\eta_{\mu\nu}h(x)\right]=-\kappa T_{\mu\nu}(x)\]
tendo como fonte uma massa pontual localizada na origem do sistema de coordenadas, ou seja, $T^{\mu\nu}(x)=M\delta^\mu_0\delta^\nu_0\delta^{(3)}({\bf x})$. A transformada de Fourier deste campo é dada por
\[h_{\mu\nu}(q)=\int d^4x\;e^{iq\cdot x}h_{\mu\nu}(x)=\underbrace{\frac{\kappa M}{4{\bf q}^2}(\eta_{\mu\nu}-2\eta_{\mu 0}\eta_{\nu 0})}_{\equiv h^{ext}_{\mu\nu}({\bf q})}2\pi\delta(q_0)\]
onde $q=p'-p$ é o momento trocado cuja componente $q_0=0$ (o campo externo $h_{\mu\nu}(x)$ é estático no nosso caso, o que implica na conservação da energia), portanto $|{\bf p'}|=|{\bf p}|$ e $|{\bf q}|=2|{\bf p}|\sin\frac{\theta}{2}$ onde $\theta$ é o ângulo de espalhamento.

Assim acrescentamos a seguinte regra de Feynman ao capítulo anterior

\noindent{\bf Campo externo}
\begin{center}
\begin{picture}(250,20)(0,0)
\Line(-5,5)(5,15)
\Line(-5,15)(5,5)
\Gluon(0,10)(60,10){3}{6}
\LongArrow(25,17)(35,17)
\Text(60,0)[r]{$\alpha\beta$}
\Text(30,0)[c]{$q$}
\Text(80,10)[l]{\large $h_{\alpha\beta}^{ext}({\bf q})=\frac{\kappa M}{4{\bf q}^2}(\eta_{\alpha\beta}-2\delta_\alpha^0\delta_\beta^0)$}
\end{picture}
\end{center}
onde fica subentendida a conservação de energia no vértice $(q_0=0)$.

Nas seções seguintes trataremos diversos casos de espalhamento para partículas com $s=0,\frac{1}{2},1,2$ e $m=0$ ou $m\neq 0$. Na seção final vamos enunciar um teorema sobre as seções de choque.
\section{Klein-Gordon}
O cálculo para a partícula escalar é o mais simples de todos devido à ausência de graus de liberdade internos. A amplitude de Feynman $\mathcal{M}$ para o processo em questão é dada por
\begin{center}
\begin{picture}(90,104)(0,0)
\Line(-5,47)(5,57)
\Line(-5,57)(5,47)
\Gluon(0,52)(60,52){3}{6}
\LongArrow(25,59)(35,59)
\Text(60,42)[r]{$\alpha\beta$}
\Text(30,42)[c]{$q$}
\DashArrowLine(60,52)(90,104){5}
\Text(80,78)[l]{$p'=p+q$}
\DashArrowLine(90,0)(60,52){5}
\Text(80,26)[l]{$p$}
\end{picture}
\end{center}
\begin{eqnarray*}
i\mathcal{M}&=&ih^{ext}_{\alpha\beta}({\bf q})V^{\alpha\beta}(p',p)\\&=&\frac{4\pi GM}{{\bf q}^2}\underbrace{(\eta_{\alpha\beta}-2\delta_\alpha^0\delta_\beta^0)[p'^\alpha p^\beta+p'^\beta p^\alpha-\eta^{\alpha\beta}(p'\cdot p-m^2)]}_{-(2m^2+4{\bf p}^2)}\\&=&-\frac{4\pi GM}{{\bf q}^2}(2m^2+4{\bf p}^2).
\end{eqnarray*}
Por conveniência, nós multiplicamos por $i$ a amplitude $\mathcal{M}$ para ficarmos com uma expressão real. A partir desta expressão, podemos calcular a energia potencial gravitacional (ou simplesmente o potencial) definida como \cite{Gupta66,Iwasaki,Gupta80}
\begin{eqnarray*}
V(r)&=&\left.\int\frac{d^3q}{(2\pi)^3}e^{i{\bf q}\cdot{\bf r}}\frac{i\mathcal{M}}{\sqrt{2E'}\sqrt{2E}}\right|_{E'=E}\\&=&-4\pi GM\left(\frac{m^2+2{\bf p}^2}{E}\right)\underbrace{\int\frac{d^3q}{(2\pi)^3}\frac{e^{i{\bf q}\cdot{\bf r}}}{{\bf q}^2}}_{\frac{1}{4\pi r}}=-\frac{GM}{r}\left(\frac{m^2+2{\bf p}^2}{E}\right)
\end{eqnarray*}
que se reduz no limite não-relativístico ao bem conhecido potencial Newtoniano $V(r)=-\frac{GMm}{r}$ e a $V(r)=-\frac{2GME}{r}$ no caso ultra-relativístico (ou equivalente para $m=0$).

Já a seção de choque diferencial é dada por \cite{Golowich,Accioly02}
\[\frac{d\sigma}{d\Omega}=\frac{|i\mathcal{M}|^2}{(4\pi)^2}=\left(\frac{GM}{\sin^2\frac{\theta}{2}}\right)^2\left(1+\frac{\alpha}{2}\right)^2\]
onde $\alpha\equiv\frac{m^2}{{\bf p}^2}=\frac{1-{\bf v}^2}{{\bf v}^2}$. Desta expressão podemos obter o ângulo de espalhamento $\theta$ através da aproximação clássica \cite{Landau} para a seção de choque $\frac{d\sigma}{d\Omega}=\frac{b}{\sin\theta}\left|\frac{db}{d\theta}\right|$, onde $b$ é o parâmetro de impacto. Assim temos \[\theta=2\sin^{-1}\left[\frac{2GM}{b}\left(1+\frac{\alpha}{2}\right)\right]\]
que para valores pequenos de $\frac{GM}{b}$ (o valor máximo de $\frac{GM}{b}$ para o Sol, por exemplo, é da ordem de $10^{-6}$, o que nos permite fazer tal aproximação) pode ser aproximado para
\[\theta\approx\frac{4GM}{b}\left(1+\frac{\alpha}{2}\right)\]
que se reduz ao valor $\theta_{Newton}=\frac{2GM}{b{\bf v}^2}$ para $|{\bf v}|\ll 1$ e a $\theta_{Einstein}=\frac{4GM}{b}$ quando $|{\bf v}|=1$ (no exemplo do Sol $\theta_{Einstein}=1.75\;arcsec$).
\section{Dirac}
O espalhamento de elétrons pelo campo gravitacional é bem mais complicado do que o caso anterior devido ao spin. O diagrama abaixo representa o espalhamento em que a partícula inicial está especificada pelo spinor $u_r({\bf p})$ e o estado final pelo spinor $\bar{u}_{r'}({\bf p'})$,
\begin{center}
\begin{picture}(90,104)(0,0)
\Line(-5,47)(5,57)
\Line(-5,57)(5,47)
\Gluon(0,52)(60,52){3}{6}
\LongArrow(25,59)(35,59)
\Text(60,42)[r]{$\alpha\beta$}
\Text(30,42)[c]{$q$}
\ArrowLine(60,52)(90,104)
\Text(90,96)[l]{$r'$}
\Text(80,78)[l]{$p'=p+q$}
\ArrowLine(90,0)(60,52)
\Text(90,8)[l]{$r$}
\Text(80,26)[l]{$p$}
\end{picture}
\end{center}
\begin{eqnarray*}
i\mathcal{M}_{r',r}&=&ih^{ext}_{\alpha\beta}({\bf q})\bar{u}_{r'}({\bf p'})V^{\alpha\beta}(p',p)u_r({\bf p})\\&=&-\frac{4\pi GME}{{\bf q}^2}\bar{u}_{r'}({\bf p'})\left(2\gamma^0-\frac{m}{E}\right)u_r({\bf p})
\end{eqnarray*}
(o potencial pode ser obtido através da expressão acima \cite{Gupta66}, porém este difere do potencial da seção anterior apenas por termos que só contribuem em distâncias da ordem do comprimento de onda Compton $\lambda=\frac{1}{m}$ e, portanto, podem ser desprezados em grandes distâncias).

Para a seção de choque não polarizada temos \cite{Papini,Acciolym}
\[\frac{d\sigma}{d\Omega}=\frac{1}{(4\pi)^2}\frac{1}{2}\sum_{r',r}|i\mathcal{M}_{r',r}|^2(2m)^2=\left(\frac{GM}{\sin^2\frac{\theta}{2}}\right)^2\left[\cos^2\frac{\theta}{2}+\frac{\alpha}{4}\left(1+\alpha+3\cos^2\frac{\theta}{2}\right)\right]\]
onde usamos $\displaystyle\sum_{r=1}^2u_r({\bf p})\bar{u}_{r}({\bf p})=\frac{\not p+m}{2m}$.
\section{Maxwell}
A amplitude $\mathcal{M}$ para o espalhamento de um fóton é dada por
\begin{center}
\begin{picture}(120,120)(0,0)
\Line(-5,55)(5,65)
\Line(-5,65)(5,55)
\Gluon(0,60)(60,60){-3}{6}
\LongArrow(25,67)(35,67)
\Text(30,50)[c]{$q$}
\Text(56,50)[r]{$\alpha\beta$}
\Photon(90,111.96)(60,60){3}{6}
\LongArrow(62.5,81.65)(66.5,90.31)
\Text(90,85.98)[l]{$p'=p+q$}
\Text(82,105.96)[rb]{$\nu$}
\Photon(90,8.04)(60,60){3}{6}
\LongArrow(67.5,29.69)(62.5,38.35)
\Text(90,34.02)[l]{$p$}
\Text(82,14.04)[rt]{$\mu$}
\end{picture}
\end{center}
\begin{eqnarray*}
i\mathcal{M}_{r',r}&=&ih_{ext}^{\alpha\beta}({\bf q})\epsilon^\nu_{r'}({\bf p'})V_{\alpha\beta,\mu\nu}(p',p)\epsilon^\mu_r({\bf p})\\&=&-\frac{8\pi GM}{{\bf q}^2}\epsilon^\nu_{r'}({\bf p'})[(\eta_{\mu\nu}-2\delta_\mu^0\delta_\nu^0)p'\cdot p-p'_\mu p_\nu\\&&+2Ep'_\mu\delta_\nu^0+2Ep_\nu\delta_\mu^0-2E^2\eta_{\mu\nu}]\epsilon^\mu_r({\bf p})
\end{eqnarray*}
onde $\epsilon^\mu_r({\bf p})$ e $\epsilon^\nu_{r'}({\bf p'})$ são as polarizações inicial e final do fóton, respectivamente, e usando $\displaystyle\Pi^{\mu\nu}({\bf p})\equiv\sum_{r=1}^2\epsilon^\mu_r({\bf p})\epsilon^\nu_r({\bf p})=-\eta^{\mu\nu}-\frac{1}{(p\cdot n)^2}[p^\mu p^\nu-p\cdot n(p^\mu n^\nu+p^\nu n^\mu)]$ onde $n^2=1$, obtemos \cite{Papini,Golowich}
\[\frac{d\sigma}{d\Omega}=\frac{1}{(4\pi)^2}\frac{1}{2}\sum_{r',r}|i\mathcal{M}_{r',r}|^2=\left(\frac{GM}{\sin^2\frac{\theta}{2}}\right)^2\cos^4\frac{\theta}{2}.\]
\section{Proca}
De modo análogo temos o caso do ``fóton massivo"
\begin{center}
\begin{picture}(120,120)(0,0)
\Line(-5,55)(5,65)
\Line(-5,65)(5,55)
\Gluon(0,60)(60,60){-3}{6}
\LongArrow(25,67)(35,67)
\Text(30,50)[c]{$q$}
\Text(56,50)[r]{$\alpha\beta$}
\Photon(90,111.96)(60,60){3}{6}
\LongArrow(62.5,81.65)(66.5,90.31)
\Text(90,85.98)[l]{$p'=p+q$}
\Text(82,105.96)[rb]{$\nu$}
\Photon(90,8.04)(60,60){3}{6}
\LongArrow(67.5,29.69)(62.5,38.35)
\Text(90,34.02)[l]{$p$}
\Text(82,14.04)[rt]{$\mu$}
\end{picture}
\end{center}
\begin{eqnarray*}
i\mathcal{M}_{r',r}&=&ih_{ext}^{\alpha\beta}({\bf q})\epsilon^\nu_{r'}({\bf p'})V_{\alpha\beta,\mu\nu}(p',p)\epsilon^\mu_r({\bf p})\\&=&-\frac{8\pi GM}{{\bf q}^2}\epsilon^\nu_{r'}({\bf p'})[(\eta_{\mu\nu}-2\delta_\mu^0\delta_\nu^0)(p'\cdot p-m^2)-p'_\mu p_\nu\\&&+2Ep'_\mu\delta_\nu^0+2Ep_\nu\delta_\mu^0-2E^2\eta_{\mu\nu}+m^2\eta_{\mu\nu}]\epsilon^\mu_r({\bf p})
\end{eqnarray*}
onde $\epsilon^\mu_r({\bf p})$ e $\epsilon^\nu_{r'}({\bf p'})$ são as polarizações inicial e final do ``fóton massivo", respectivamente, e como $\displaystyle\sum_{r=1}^3\epsilon^\mu_r({\bf p})\epsilon^\nu_r({\bf p})=-\eta^{\mu\nu}+\frac{p^\mu p^\nu}{m^2}$, obtemos \cite{Paszko}
\[\frac{d\sigma}{d\Omega}=\frac{1}{(4\pi)^2}\frac{1}{3}\sum_{r',r}|i\mathcal{M}_{r',r}|^2=\left(\frac{GM}{\sin^2\frac{\theta}{2}}\right)^2\left[\frac{1}{3}+\frac{2}{3}\cos^4\frac{\theta}{2}-\frac{\alpha}{3}\left(1-\frac{3\alpha}{4}-4\cos^2\frac{\theta}{2}\right)\right].\]
\section{Einstein}
Vamos considerar também o espalhamento de um gráviton pelo campo externo
\begin{center}
\begin{picture}(120,120)(0,0)
\Line(-5,55)(5,65)
\Line(-5,65)(5,55)
\Gluon(0,60)(60,60){3}{6}
\LongArrow(35,67)(25,67)
\Text(30,50)[c]{$q$}
\Text(56,50)[r]{$\mu\nu$}
\Gluon(90,111.96)(60,60){3}{6}
\LongArrow(62.5,81.65)(66.5,90.31)
\Text(90,85.98)[l]{$p'=p-q$}
\Text(82,105.96)[rb]{$\lambda\rho$}
\Gluon(90,8.04)(60,60){3}{6}
\LongArrow(67.5,29.69)(62.5,38.35)
\Text(90,34.02)[l]{$p$}
\Text(82,14.04)[rt]{$\alpha\beta$}
\end{picture}
\end{center}
\[i\mathcal{M}_{r',r}=ih_{ext}^{\mu\nu}({\bf q})\epsilon^{\lambda\rho}_{r'}({\bf p'})V_{\alpha\beta,\mu\nu,\lambda\rho}(p,q,p')\epsilon^{\alpha\beta}_r({\bf p})\]
onde a soma de polarizações é dada por \cite{Weinberg64}
\[\displaystyle\sum_{r=1}^2\epsilon^{\alpha\beta}_r({\bf p})\epsilon^{\mu\nu}_r({\bf p})=\frac{1}{2}\left[\Pi^{\alpha\mu}({\bf p})\Pi^{\beta\nu}({\bf p})+\Pi^{\alpha\nu}({\bf p})\Pi^{\beta\mu}({\bf p})-\Pi^{\alpha\beta}({\bf p})\Pi^{\mu\nu}({\bf p})\right]\]
assim
\[\frac{d\sigma}{d\Omega}=\frac{1}{(4\pi)^2}\frac{1}{2}\sum_{r',r}|i\mathcal{M}_{r',r}|^2=\left(\frac{GM}{\sin^2\frac{\theta}{2}}\right)^2\left(\sin^8\frac{\theta}{2}+\cos^8\frac{\theta}{2}\right).\]
\section{Teorema}
Na Tabela 2.1 estão agrupadas as seções de choque obtidas neste capítulo, vemos então o seguinte:
\begin{center}
{\it As seções de choque são distintas para partículas de diferentes spins $s$.}
\end{center}

Portanto existe, já em $1^a$ ordem no campo externo, uma violação do Princípio da Equivalência \cite{Accioly1}.
\begin{table}[h]
\begin{center}
\begin{tabular}{|c|c|c|}
\hline 
m & s & $d\sigma/d\Omega$\\
\hline 
0&0&$\left(\frac{GM}{\sin^2\frac{\theta}{2}}\right)^2$\\
$\neq 0$&0&$\left(\frac{GM}{\sin^2\frac{\theta}{2}}\right)^2\left(1+\frac{\alpha}{2}\right)^2$\\
0&1/2&$\left(\frac{GM}{\sin^2\frac{\theta}{2}}\right)^2\cos^2\frac{\theta}{2}$\\
$\neq 0$&1/2&$\left(\frac{GM}{\sin^2\frac{\theta}{2}}\right)^2\left[\cos^2\frac{\theta}{2}+\frac{\alpha}{4}\left(1+\alpha+3\cos^2\frac{\theta}{2}\right)\right]$\\
0&1&$\left(\frac{GM}{\sin^2\frac{\theta}{2}}\right)^2\cos^4\frac{\theta}{2}$\\
$\neq 0$&1&$\left(\frac{GM}{\sin^2\frac{\theta}{2}}\right)^2\left[\frac{1}{3}+\frac{2}{3}\cos^4\frac{\theta}{2}-\frac{\alpha}{3}\left(1-\frac{3\alpha}{4}-4\cos^2\frac{\theta}{2}\right)\right]$\\
0&2&$\left(\frac{GM}{\sin^2\frac{\theta}{2}}\right)^2\left(\sin^8\frac{\theta}{2}+\cos^8\frac{\theta}{2}\right)$\\
\hline 
\end{tabular}
\caption{Seções de Choque Diferenciais (não polarizadas).}
\end{center}
\end{table}
\newpage
No entanto, a diferença entre estas expressões é muito pequena. Na Figura 2.1 temos o gráfico de
\[\frac{\Delta(d\sigma/d\Omega)}{(d\sigma/d\Omega)_{s=0}}=\frac{(d\sigma/d\Omega)_s-(d\sigma/d\Omega)_{s=0}}{(d\sigma/d\Omega)_{s=0}}\]
em função de $\theta$ para partículas de $m=0$.
\begin{figure}[h]
\begin{center}
\centering
\includegraphics[angle=-90,scale=.59]{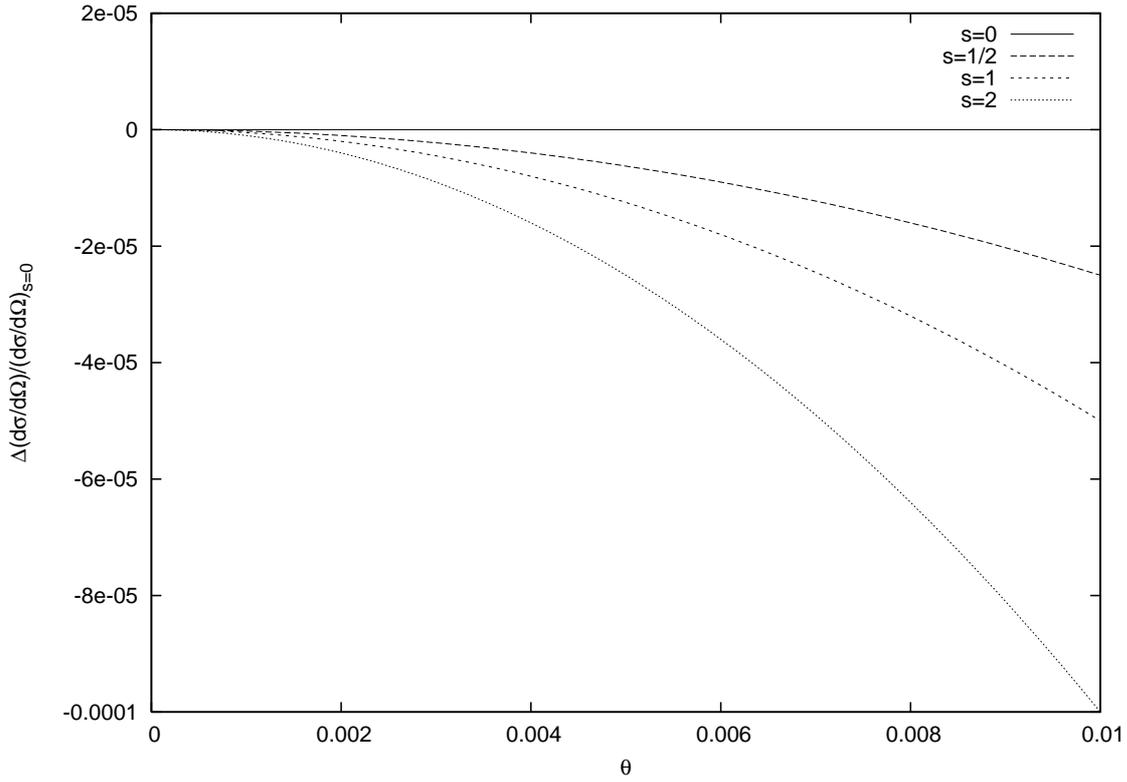}
\end{center}
\caption{\small $\Delta(d\sigma/d\Omega)/(d\sigma/d\Omega)_{s=0}$ em função de $\theta$.}
\end{figure}

Vemos que para ângulos pequenos $\Delta(d\sigma/d\Omega)/(d\sigma/d\Omega)_{s=0}\approx-s\theta^2/2$ e para valores típicos do ângulo de deflexão $\theta\sim 10^{-6}$ temos $\Delta(d\sigma/d\Omega)/(d\sigma/d\Omega)_{s=0}\sim 10^{-12}$ que é muito pequeno.

\chapter{Campo Externo - $2^a$ ordem}
Como um aquecimento para o caso do fóton, vamos começar calculando as correções de $2^a$ ordem no campo externo para uma partícula escalar \cite{Uno,Accioly2}.

Por estarmos considerando que o campo é fraco porém a massa $M$ é ainda muito maior do que a energia da partícula espalhada (tipicamente para o caso de um fóton, digamos, na faixa do microondas $\nu\sim 1GHz$, espalhado pelo campo gravitacional do Sol temos $\frac{E}{M}\sim 10^{-71}$) nós podemos desprezar o Bremsstrahlung e as correções radiativas, pois esses efeitos estão na razão de $\frac{E}{M}$ em relação aos gráficos de $2^a$ ordem no campo externo.
\section{Escalar}
Lembrando o resultado do capítulo anterior
\begin{center}
\begin{picture}(90,104)(0,0)
\Line(-5,47)(5,57)
\Line(-5,57)(5,47)
\Gluon(0,52)(60,52){3}{6}
\LongArrow(25,59)(35,59)
\Text(60,42)[r]{$\alpha\beta$}
\Text(30,42)[c]{$q$}
\DashArrowLine(60,52)(90,104){5}
\Text(80,78)[l]{$p'=p+q$}
\DashArrowLine(90,0)(60,52){5}
\Text(80,26)[l]{$p$}
\end{picture}
\end{center}
\[i\mathcal{M}^{(1)}=-\frac{4\pi GM}{{\bf q}^2}(2m^2+4{\bf p}^2)\]
e que existem somente 3 gráficos que contribuem para $2^a$ ordem,
\begin{center}
\begin{picture}(84,94)(0,0)
\Line(-5,-5)(5,5)
\Line(-5,5)(5,-5)
\Gluon(0,0)(42,42){3}{6}
\LongArrow(30.5,17.5)(38.5,24.5)
\Text(9,21)[r]{$k$}
\Text(14,2)[l]{$\alpha\beta$}
\DashArrowLine(84,0)(42,42){5}
\Text(75,21)[l]{$p$}
\Line(-5,89)(5,79)
\Line(-5,79)(5,89)
\Gluon(42,42)(0,84){-3}{6}
\LongArrow(38.5,59.5)(30.5,66.5)
\Text(9,63)[r]{$k-q$}
\Text(14,81)[l]{$\mu\nu$}
\DashArrowLine(42,42)(84,84){5}
\Text(75,63)[l]{$p'=p+q$}
\end{picture}
\end{center}
\begin{eqnarray*}
i\mathcal{M}^{(2)}&=&\frac{i}{2}\int\frac{d^3k}{(2\pi)^3}h_{\alpha\beta}^{ext}({\bf k})h_{\mu\nu}^{ext}({\bf k-q})V^{\mu\nu,\alpha\beta}(p',p)\\&=&-16\pi^2G^2M^2(8E^2-{\bf q}^2)\underbrace{\int\frac{d^3k}{(2\pi)^3}\frac{1}{{\bf k}^2({\bf k-q})^2}}_{\frac{1}{8|{\bf q}|}}\\&=&-\frac{4\pi^2G^2M^2}{|{\bf q}|}\left(4E^2-\frac{{\bf q}^2}{2}\right),
\end{eqnarray*}
onde usamos
\[(\eta_{\alpha\beta}-2\delta_\alpha^0\delta_\beta^0)V^{\alpha\beta,\mu\nu}(p',p)(\eta_{\mu\nu}-2\delta_\mu^0\delta_\nu^0)=\frac{i\kappa^2}{2}(8E^2-{\bf q}^2).\]

Note que existe um fator de simetria $\frac{1}{2}$ nos gráficos de $\mathcal{M}^{(2)}$ e $\mathcal{M}^{(3)}$
\begin{center}
\begin{picture}(150,122)(0,0)
\Line(25,3.04)(35,13.04)
\Line(25,13.04)(35,3.04)
\Line(25,106.96)(35,116.96)
\Line(25,116.96)(35,106.96)
\Gluon(60,60)(120,60){-3}{6}
\LongArrow(85,67)(95,67)
\Text(90,50)[c]{$q$}
\Text(60,49)[l]{$\mu\nu$}
\Text(120,50)[r]{$\sigma\tau$}
\Gluon(30,111.96)(60,60){3}{6}
\LongArrow(59.5,81.65)(55.5,90.31)
\Text(35,85.98)[r]{$k-q$}
\Text(36,105.96)[lb]{$\lambda\rho$}
\Gluon(30,8.04)(60,60){3}{6}
\LongArrow(52.5,29.69)(57.5,38.35)
\Text(35,34.02)[r]{$k$}
\Text(36,14.04)[lt]{$\alpha\beta$}
\DashArrowLine(120,60)(150,111.96){5}
\Text(142,85.98)[l]{$p'=p+q$}
\DashArrowLine(150,8.04)(120,60){5}
\Text(142,34.02)[l]{$p$}
\end{picture}
\end{center}
\begin{eqnarray*}
i\mathcal{M}^{(3)}&=&\frac{i}{2}\int^\Lambda\frac{d^3k}{(2\pi)^3}h_{\alpha\beta}^{ext}({\bf k})h_{\lambda\rho}^{ext}({\bf k-q})V^{\alpha\beta,\mu\nu,\lambda\rho}(k,q,k-q)\frac{i\mathcal{P}_{\mu\nu,\sigma\tau}}{q^2}V^{\sigma\tau}(p',p)\\&=&-\frac{4\pi^2G^2M^2}{|{\bf q}|}\left(-m^2-\frac{{\bf p}^2}{4}+\frac{7{\bf q}^2}{16}\right)+\mathcal{O}(\Lambda).
\end{eqnarray*}
Neste segundo gráfico precisamos da seguinte expressão
\begin{eqnarray*}
(\eta_{\alpha\beta}-2\delta_\alpha^0\delta_\beta^0)V^{\alpha\beta,\mu\nu,\lambda\rho}(k,q,k-q)(\eta_{\lambda\rho}-2\delta_\lambda^0\delta_\rho^0)&=&i\kappa[2(2k\cdot q-q^2-2k^2)\delta^\mu_0\delta^\nu_0\\&&+2q^\mu q^\nu-2k^\mu k^\nu+q^\mu k^\nu+q^\nu k^\mu\\&&+(k^2-k\cdot q)\eta^{\mu\nu}]
\end{eqnarray*}
e portanto das seguintes integrais
\begin{eqnarray*}
\int\frac{d^3k}{(2\pi)^3}\frac{k^i}{{\bf k}^2({\bf k-q})^2}&=&\frac{q^i}{16|{\bf q}|}\\
\int^\Lambda\frac{d^3k}{(2\pi)^3}\frac{k^ik^j}{{\bf k}^2({\bf k-q})^2}&\approx&\frac{|{\bf q}|}{64}\left(3\frac{q^iq^j}{{\bf q}^2}-\delta^{ij}\right)+\frac{\Lambda\delta^{ij}}{6\pi^2}.
\end{eqnarray*}
Note que a segunda integral tem divergência linear no ultravoleta e por isso introduzimos um cut-off $\Lambda$.

No entanto, essa divergência é espúria, pois o mesmo resultado pode ser obtido diretamente através da aproximação de $2^a$ ordem no {\it próprio} campo externo.

Por exemplo, em coordenadas isotrópicas a métrica é dada por \cite{Weinberg}
\begin{eqnarray*}
g_{00}&=&\left(\frac{1-\frac{GM}{2r}}{1+\frac{GM}{2r}}\right)^2=1-\frac{2GM}{r}+\frac{2G^2M^2}{r^2}+\ldots\\
g_{0i}&=&0\\
g_{ij}&=&-\left(1+\frac{GM}{2r}\right)^4\delta_{ij}=-\left(1+\frac{2GM}{r}+\frac{3G^2M^2}{2r^2}+\ldots\right)\delta_{ij}
\end{eqnarray*}
e assim
\begin{eqnarray*}
\kappa h^{extra}_{00}({\bf q})&=&2G^2M^2\underbrace{\int d^3x\frac{e^{-i{\bf q}\cdot {\bf x}}}{r^2}}_{\frac{2\pi^2}{|{\bf q}|}}=\frac{4\pi^2G^2M^2}{|{\bf q}|}\\
\kappa h^{extra}_{0i}({\bf q})&=&0\\
\kappa h^{extra}_{ij}({\bf q})&=&-\frac{3G^2M^2}{2}\delta_{ij}\int d^3x\frac{e^{-i{\bf q}\cdot {\bf x}}}{r^2}=-\frac{3\pi^2G^2M^2}{|{\bf q}|}\delta_{ij}
\end{eqnarray*}
obtemos o mesmo resultado sem a divergência
\[i\mathcal{M}^{(1)}_{extra}=ih^{extra}_{\mu\nu}({\bf q})V^{\mu\nu}(p',p)=-\frac{4\pi^2G^2M^2}{|{\bf q}|}\left(-m^2-\frac{{\bf p}^2}{4}+\frac{7{\bf q}^2}{16}\right).\]
\begin{center}
\begin{picture}(90,164)(0,-60)
\Line(-5,47)(5,57)
\Line(-5,57)(5,47)
\Gluon(0,52)(60,52){3}{6}
\LongArrow(25,59)(35,59)
\Text(60,62)[r]{$\mu\nu$}
\Text(30,42)[c]{$p'-k$}
\DashArrowLine(60,52)(90,104){5}
\Text(80,78)[l]{$p'=p+q$}
\DashArrowLine(60,-8)(60,52){5}
\Text(64,22)[l]{$k$}
\DashArrowLine(90,-60)(60,-8){5}
\Text(80,-34)[l]{$p$}
\Line(-5,-13)(5,-3)
\Line(-5,-3)(5,-13)
\Gluon(0,-8)(60,-8){3}{6}
\LongArrow(25,-15)(35,-15)
\Text(60,-18)[r]{$\alpha\beta$}
\Text(30,2)[c]{$k-p$}
\end{picture}
\end{center}
esse gráfico é divergente no infravermelho, por isso temos que usar uma massa $\mu$ para o gráviton. Veja o Apêndice A para o cálculo de $I$. Em especial só precisamos do resultado para $\mu\rightarrow 0$
\begin{eqnarray*}
i\mathcal{M}^{(4)}&=&i\int\frac{d^3k}{(2\pi)^3}h_{\mu\nu}^{ext}({\bf p'-k})V^{\mu\nu}(p',k)\frac{i}{k^2-m^2+i\epsilon}V^{\alpha\beta}(k,p)h_{\alpha\beta}^{ext}({\bf k-p})\\&=&\frac{2G^2M^2}{\pi}(2m^2+4{\bf p}^2)^2\underbrace{\int\frac{d^3k}{[({\bf p'}-{\bf k})^2+\mu^2][({\bf p}-{\bf k})^2+\mu^2]({\bf p}^2-{\bf k}^2+i\epsilon)}}_{I\approx-i\frac{2\pi^2}{|{\bf p}|{\bf q}^2}\ln\frac{|{\bf q}|}{\mu}}\\&\approx&-i\frac{4\pi G^2M^2}{|{\bf p}|{\bf q}^2}(2m^2+4{\bf p}^2)^2\ln \frac{|{\bf q}|}{\mu}
\end{eqnarray*}
onde usamos
\[(\eta_{\alpha\beta}-2\delta_\alpha^0\delta_\beta^0)[k^\alpha p^\beta+k^\beta p^\alpha-\eta^{\alpha\beta}(k\cdot p-m^2)]=2m^2-4Ek_0=2m^2-4E^2=-(2m^2+4{\bf p}^2)\]
pois $k_0=E$ pela conservação de energia nos vértices de campo externo, e portanto
\[k^2-m^2+i\epsilon={\bf p}^2-{\bf k}^2+i\epsilon.\]

Assim conseguimos calcular a seção de choque diferencial em $2^a$ ordem \cite{Uno,Accioly2}
\[\frac{d\sigma}{d\Omega}=\frac{|i\mathcal{M}^{total}|^2}{(4\pi)^2}=\left(\frac{d\sigma}{d\Omega}\right)_0\left[1+\frac{\pi GM|{\bf p}|\sin\frac{\theta}{2}}{1+\frac{\alpha}{2}}\left(3\alpha+\frac{15-\sin^2\frac{\theta}{2}}{4}\right)\right]\]
onde $\displaystyle\mathcal{M}^{total}=\sum_{i=1}^4\mathcal{M}^{(i)}$ e \[\left(\frac{d\sigma}{d\Omega}\right)_0=\left(\frac{GM}{\sin^2\frac{\theta}{2}}\right)^2\left(1+\frac{\alpha}{2}\right)^2\]
é a seção de choque em $1^a$ ordem no campo externo. Observe que a divergência infravermelha não contribui na ordem de aproximação calculada. De fato pode-se mostrar que há um cancelamento destas divergências na gravitação \cite{Weinberg65,Gupta69}.
\section{Fóton}
Do capítulo anterior
\begin{center}
\begin{picture}(120,120)(0,0)
\Line(-5,55)(5,65)
\Line(-5,65)(5,55)
\Gluon(0,60)(60,60){-3}{6}
\LongArrow(25,67)(35,67)
\Text(30,50)[c]{$q$}
\Text(56,50)[r]{$\alpha\beta$}
\Photon(90,111.96)(60,60){3}{6}
\LongArrow(62.5,81.65)(66.5,90.31)
\Text(90,85.98)[l]{$p'=p+q$}
\Text(82,105.96)[rb]{$\nu$}
\Photon(90,8.04)(60,60){3}{6}
\LongArrow(67.5,29.69)(62.5,38.35)
\Text(90,34.02)[l]{$p$}
\Text(82,14.04)[rt]{$\mu$}
\end{picture}
\end{center}
\[i\mathcal{M}_{\mu\nu}^{(1)}=-\frac{8\pi GM}{{\bf q}^2}[(\eta_{\mu\nu}-2\delta_\mu^0\delta_\nu^0)p'\cdot p-p'_\mu p_\nu+2Ep'_\mu\delta_\nu^0+2Ep_\nu\delta_\mu^0-2E^2\eta_{\mu\nu}]\]
e novamente temos 3 gráficos em $2^a$ ordem
\begin{center}
\begin{picture}(84,89)(0,0)
\Line(-5,-5)(5,5)
\Line(-5,5)(5,-5)
\Gluon(0,0)(42,42){3}{6}
\LongArrow(30.5,17.5)(38.5,24.5)
\Text(9,21)[r]{$k$}
\Text(14,2)[l]{$\alpha\beta$}
\Photon(84,0)(42,42){3}{6}
\LongArrow(53.5,17.5)(45.5,24.5)
\Text(75,21)[l]{$p$}
\Text(64,2)[l]{$\lambda$}
\Line(-5,79)(5,89)
\Line(-5,89)(5,79)
\Gluon(42,42)(0,84){-3}{6}
\LongArrow(38.5,59.5)(30.5,66.5)
\Text(9,63)[r]{$k-q$}
\Text(14,82)[l]{$\mu\nu$}
\Photon(42,42)(84,84){3}{6}
\LongArrow(45.5,59.5)(53.5,66.5)
\Text(75,63)[l]{$p'=p+q$}
\Text(64,82)[l]{$\rho$}
\end{picture}
\end{center}
\begin{eqnarray*}
i\mathcal{M}_{\lambda\rho}^{(2)}&=&\frac{i}{2}\int\frac{d^3k}{(2\pi)^3}h^{\mu\nu}_{ext}({\bf k-q})h^{\alpha\beta}_{ext}({\bf k})V_{\alpha\beta,\mu\nu,\lambda\rho}(p',p)\\&=&\frac{i\kappa^2M^2}{32}\underbrace{(\eta^{\alpha\beta}-2\delta^\alpha_0\delta^\beta_0)(\eta^{\mu\nu}-2\delta^\mu_0\delta^\nu_0)V_{\alpha\beta,\mu\nu,\lambda\rho}(p',p)}_{-i\kappa^2(\eta_{\lambda\rho}p'\cdot p-p'_\lambda p_\rho)}\underbrace{\int\frac{d^3k}{(2\pi)^3}\frac{1}{{\bf k}^2({\bf k-q})^2}}_{\frac{1}{8|{\bf q}|}}\\&=&\frac{4\pi^2G^2M^2}{|{\bf q}|}(\eta_{\lambda\rho}p'\cdot p-p'_\lambda p_\rho)
\end{eqnarray*}
\begin{center}
\begin{picture}(150,122)(0,0)
\Line(25,3.04)(35,13.04)
\Line(25,13.04)(35,3.04)
\Line(25,106.96)(35,116.96)
\Line(25,116.96)(35,106.96)
\Gluon(60,60)(120,60){-3}{6}
\LongArrow(85,67)(95,67)
\Text(90,50)[c]{$q$}
\Text(60,49)[l]{$\mu\nu$}
\Text(120,50)[r]{$\sigma\tau$}
\Gluon(30,111.96)(60,60){3}{6}
\LongArrow(59.5,81.65)(55.5,90.31)
\Text(35,85.98)[r]{$k-q$}
\Text(36,105.96)[lb]{$\lambda\rho$}
\Gluon(30,8.04)(60,60){3}{6}
\LongArrow(52.5,29.69)(57.5,38.35)
\Text(35,34.02)[r]{$k$}
\Text(36,14.04)[lt]{$\alpha\beta$}
\Photon(120,60)(150,111.96){3}{6}
\LongArrow(120.5,81.65)(124.5,90.31)
\Text(142,85.98)[l]{$p'=p+q$}
\Text(140,105.96)[rb]{$\zeta$}
\Photon(150,8.04)(120,60){3}{6}
\LongArrow(127.5,29.69)(122.5,38.35)
\Text(142,34.02)[l]{$p$}
\Text(140,10.04)[rt]{$\epsilon$}
\end{picture}
\end{center}
\begin{eqnarray*}
i\mathcal{M}^{(3)}_{\epsilon\zeta}&=&\frac{i}{2}\int\frac{d^3k}{(2\pi)^3}h_{ext}^{\lambda\rho}({\bf k-q})h_{ext}^{\alpha\beta}({\bf k})V_{\alpha\beta,\mu\nu,\lambda\rho}(k,q,k-q)\frac{i\mathcal{P}^{\mu\nu,\sigma\tau}}{q^2}V_{\sigma\tau,\epsilon\zeta}(p',p)\\&=&-\frac{\kappa^2M^2}{32q^2}\underbrace{\mathcal{P}^{\mu\nu,\sigma\tau}V_{\sigma\tau,\epsilon\zeta}(p',p)}_{V^{\mu\nu}_{\phantom{\mu\nu}\epsilon\zeta}(p',p)}\\&&\times\underbrace{\int\frac{d^3k}{(2\pi)^3}\frac{(\eta^{\alpha\beta}-2\delta^\alpha_0\delta^\beta_0)V_{\alpha\beta,\mu\nu,\lambda\rho}(k,q,k-q)(\eta^{\lambda\rho}-2\delta^\lambda_0\delta^\rho_0)}{{\bf k}^2({\bf k-q})^2}}_{\frac{i\kappa}{32|{\bf q}|}(9q_\mu q_\nu-q^2\eta_{\mu\nu}-q^2\delta_\mu^0\delta_\nu^0)}\\&=&\frac{\pi^2G^2M^2}{2|{\bf q}|}[9p'_\zeta p_\epsilon+(\eta_{\epsilon\zeta}-2\delta_\epsilon^0\delta_\zeta^0)p'\cdot p-p'_\epsilon p_\zeta+2Ep'_\epsilon\delta_\zeta^0+2Ep_\zeta\delta_\epsilon^0-2E^2\eta_{\epsilon\zeta}]
\end{eqnarray*}
\begin{center}
\begin{picture}(90,164)(0,-60)
\Line(-5,47)(5,57)
\Line(-5,57)(5,47)
\Gluon(0,52)(60,52){3}{6}
\LongArrow(25,59)(35,59)
\Text(60,62)[r]{$\mu\nu$}
\Text(30,42)[c]{$p'-k$}
\Photon(60,52)(90,104){3}{6}
\Text(80,104)[rt]{$\tau$}
\Text(80,78)[l]{$p'=p+q$}
\LongArrow(65,73.67)(70,82.33)
\Photon(60,-8)(60,52){3}{6}
\LongArrow(53,17)(53,27)
\Text(64,22)[l]{$k$}
\Text(66,50)[lt]{$\sigma$}
\Text(66,-6)[lb]{$\rho$}
\Photon(90,-60)(60,-8){3}{6}
\Text(80,-60)[rb]{$\lambda$}
\Text(80,-34)[l]{$p$}
\LongArrow(70,-38.33)(65,-29.67)
\Line(-5,-13)(5,-3)
\Line(-5,-3)(5,-13)
\Gluon(0,-8)(60,-8){3}{6}
\LongArrow(25,-15)(35,-15)
\Text(60,-18)[r]{$\alpha\beta$}
\Text(30,2)[c]{$k-p$}
\end{picture}
\end{center}
\begin{eqnarray*}
i\mathcal{M}^{(4)}_{\lambda\tau}&=&i\int\frac{d^3k}{(2\pi)^3}h^{\mu\nu}_{ext}({\bf p'-k})V_{\mu\nu,\sigma\tau}(p',k)\frac{-i\eta^{\sigma\rho}}{k^2+i\epsilon}V_{\alpha\beta,\lambda\rho}(k,p)h^{\alpha\beta}_{ext}({\bf k-p})\\&=&\frac{GM^2}{4\pi^2}\int d^3k\frac{(\eta^{\mu\nu}-2\delta^\mu_0\delta^\nu_0)V_{\mu\nu,\sigma\tau}(p',k)\eta^{\sigma\rho}(\eta^{\alpha\beta}-2\delta^\alpha_0\delta^\beta_0)V_{\alpha\beta,\lambda\rho}(k,p)}{[({\bf p'}-{\bf k})^2+\mu^2][({\bf p}-{\bf k})^2+\mu^2]({\bf p}^2-{\bf k}^2+i\epsilon)}
\end{eqnarray*}
onde
\[(\eta^{\mu\nu}-2\delta^\mu_0\delta^\nu_0)V_{\mu\nu,\sigma\tau}(p',k)=i\kappa[(\eta_{\sigma\tau}-2\delta_\sigma^0\delta_\tau^0)p'\cdot k-p'_\sigma k_\tau+2Ep'_\sigma\delta_\tau^0+2Ek_\tau\delta_\sigma^0-2E^2\eta_{\sigma\tau}]\]
assim
\begin{eqnarray*}
&(\eta^{\mu\nu}-2\delta^\mu_0\delta^\nu_0)V_{\mu\nu,\sigma\tau}(p',k)\eta^{\sigma\rho}(\eta^{\alpha\beta}-2\delta^\alpha_0\delta^\beta_0)V_{\alpha\beta,\lambda\rho}(k,p)&\\&=-\kappa^2\{4E^2(E^2\eta_{\lambda\tau}-Ep'_\lambda\delta^0_\tau-Ep_\tau\delta^0_\lambda+p'\cdot p\delta^0_\lambda\delta^0_\tau)&\\&+k_\tau(2E^2p'_\lambda-2E\delta^0_\lambda p'\cdot p)+k_\lambda(2E^2p_\tau-2E\delta^0_\tau p'\cdot p)&\\&+k^\alpha[-2E^2\eta_{\lambda\tau}(p'_\alpha+p_\alpha)+2E(p'_\alpha\delta^0_\lambda p_\tau+p_\alpha\delta^0_\tau p'_\lambda)]&\\&+k^\alpha k^\beta(p'_\alpha p_\beta \eta_{\lambda\tau}-\eta_{\alpha\lambda}p'_\beta p_\tau-\eta_{\alpha\tau}p_\beta p'_\lambda+\eta_{\alpha\lambda}\eta_{\beta\tau}p'\cdot p)\}&
\end{eqnarray*}
e usando a parte real (só esta parte contribui) das integrais do Apêndice A (lembre que $k_0=E$ nesse diagrama)
\[(I,I^\alpha,I^{\alpha\beta})=\int\frac{(1,k^\alpha,k^\alpha k^\beta)d^3k}{[({\bf p'}-{\bf k})^2+\mu^2][({\bf p}-{\bf k})^2+\mu^2]({\bf p}^2-{\bf k}^2+i\epsilon)}\]
\begin{eqnarray*}
I&=&0\\
I^\alpha&=&\left(0,\frac{-\pi^3({\bf p'+p})^i}{8|{\bf p}|^3\sin\frac{\theta}{2}(1+\sin\frac{\theta}{2})}\right)\\
I^{\alpha\beta}&=&\left(\begin{array}{cc}0&EI^i\\EI^j&\frac{-\pi^3}{4|{\bf p}|(1+\sin\frac{\theta}{2})}\left[\delta^{ij}+\frac{{\bf q}^i{\bf q}^j}{4{\bf p}^2\sin\frac{\theta}{2}}+\frac{(2+\sin\frac{\theta}{2})({\bf p'+p})^i({\bf p'+p})^j}{4{\bf p}^2\sin\frac{\theta}{2}(1+\sin\frac{\theta}{2})}\right]\\\end{array}\right)
\end{eqnarray*}
obtemos
\begin{eqnarray*}
i\mathcal{M}^{(1)}_{\mu\nu}i\mathcal{M}^{(4)\mu\nu}&=&\frac{64G^3M^3}{{\bf q}^2}[p'\cdot p(p'\cdot p-2E^2)I^{\mu\nu}\eta_{\mu\nu}-2p'\cdot pI^{\mu\nu}p'_\mu p_\nu\\&&-2E^2(I^{\mu\nu}p_\mu p_\nu+I^{\mu\nu}p'_\mu p'_\nu)+8E^4(I^\mu p_\mu+I^\mu p'_\mu)]\\&=&\frac{16\pi^3G^3M^3E}{\sin^3\frac{\theta}{2}}\left(2+\sin\frac{\theta}{2}\right)^2\left(1-\sin\frac{\theta}{2}\right)^2
\end{eqnarray*}
além disso
\begin{eqnarray*}
i\mathcal{M}^{(1)}_{\mu\nu}i\mathcal{M}^{(1)\mu\nu}&=&\frac{32\pi^2G^2M^2}{\sin^4\frac{\theta}{2}}\cos^4\frac{\theta}{2}\\
i\mathcal{M}^{(1)}_{\mu\nu}i\mathcal{M}^{(2)\mu\nu}&=&0\\
i\mathcal{M}^{(1)}_{\mu\nu}i\mathcal{M}^{(3)\mu\nu}&=&-\frac{4\pi^3G^3M^3E}{\sin^3\frac{\theta}{2}}\cos^4\frac{\theta}{2}\\
\end{eqnarray*}
\newpage
\noindent
e finalmente \cite{Accioly3}
\[\frac{d\sigma}{d\Omega}=\left(\frac{d\sigma}{d\Omega}\right)_0\left[1+\frac{\pi GME\sin\frac{\theta}{2}}{(1+\sin\frac{\theta}{2})^2}\left(\frac{15+14\sin\frac{\theta}{2}+3\sin^2\frac{\theta}{2}}{4}\right)\right]\]
em que
\[\left(\frac{d\sigma}{d\Omega}\right)_0=\left(\frac{GM}{\sin^2\frac{\theta}{2}}\right)^2\cos^4\frac{\theta}{2}.\]

\chapter{Igualdade entre as teorias Semiclássica e Efetiva da Gravitação}
\section{Nível de Árvore}
A Teoria Efetiva da Gravitação é uma Teoria Quântica de Campos que trata o gráviton como uma partícula quantizada, mas separa as partes não analíticas $(1/q^2,1/q,\ln q)$ da amplitude de Feynman $\mathcal{M}$ dos termos analíticos $(1,q,q^2,\ldots)$, onde $q$ é o momento trocado entre as partículas, e assim sendo, válida somente em grandes distâncias (ou baixo momento trocado) \cite{Donoghue,Bjerrum}.

Neste capítulo, vamos calcular o espalhamento entre duas partículas: uma delas é um escalar de massa $M$ e momento inicial $P$ (e final $P'$). A massa $M$ vai ser considerada muito maior do que qualquer outra energia envolvida. A outra partícula pode ser em princípio de qualquer tipo, com momento inicial $p$ (e final $p'$), como no diagrama abaixo:
\begin{center}
\begin{picture}(120,103.92)(0,0)
\Gluon(30,51.96)(90,51.96){3}{6}
\LongArrow(55,61.96)(65,61.96)
\Text(60,41.96)[c]{$q$}
\Text(30,41.96)[l]{$\alpha\beta$}
\Text(90,40.96)[r]{$\mu\nu$}
\DashArrowLine(0,0)(30,51.96){5}
\Text(5,25.98)[r]{$P$}
\DashArrowLine(30,51.96)(0,103.92){5}
\Text(5,77.94)[r]{$P'=P-q$}
\ArrowLine(120,0)(90,51.96)
\Text(115,25.98)[l]{$p$}
\ArrowLine(90,51.96)(120,103.92)
\Text(115,77.94)[l]{$p'=p+q$}
\end{picture}
\end{center}

A igualdade em nível de árvore entre as teorias Semiclássica e Efetiva é trivial na aproximação $|q^0|\ll |{\bf q}|\ll M$ e no sistema de referência do laboratório, isto é, $P^\mu=(M,{\bf 0})$, pois
\begin{eqnarray*}
\mathcal{M}_{\mbox{efetiva}}&=&V^{\alpha\beta}(P',P)\frac{i\mathcal{P}_{\alpha\beta,\mu\nu}}{q^2}V^{\mu\nu}(p',p)\\&=&\underbrace{-\frac{i\kappa}{2}[P'^\alpha P^\beta+P'^\beta P^\alpha-\eta^{\alpha\beta}(P'\cdot P-M^2)]}_{\approx-i\kappa M^2\delta^\alpha_0\delta^\beta_0}\frac{i\mathcal{P}_{\alpha\beta,\mu\nu}}{q^2}V^{\mu\nu}(p',p)\\&\approx&2M\underbrace{\frac{\kappa M}{4{\bf q}^2}(\eta_{\mu\nu}-2\delta_\mu^0\delta_\nu^0)}_{h^{ext}_{\mu\nu}({\bf q})}V^{\mu\nu}(p',p)\\&=&2M\mathcal{M}_{\mbox{semiclássica}}
\end{eqnarray*}
onde $h^{ext}_{\mu\nu}({\bf q})$ é o campo externo usado nos capítulos anteriores, e portanto
\[\mathcal{M}_{\mbox{semiclássica}}\approx\frac{\mathcal{M}_{\mbox{efetiva}}}{2M}.\]
A origem do fator $\frac{1}{2M}$ é devida aos fatores de normalização $\frac{1}{\sqrt{2E'}}\frac{1}{\sqrt{2E}}\approx\frac{1}{2M}$ da partícula escalar. Em especial, para o caso em que a segunda partícula é também um escalar, temos
\[i\mathcal{M}^{(1)}\approx-\frac{8\pi GM^2}{{\bf q}^2}(2m^2+4{\bf p}^2).\]
\section{Nível de Loop}
Vamos demonstrar a igualdade entre as duas teorias também em nível de loop, para simplificar os cálculos vamos nos limitar ao caso em que as duas partículas são escalares, ainda na aproximação $|q^0|\ll |{\bf q}|\ll M$ e, além disso, vamos considerar $E\ll M$ onde $E=\sqrt{{\bf p}^2+m^2}$ é a energia da outra partícula.

Dos diversos gráficos existentes, apenas os seguintes contribuem na aproximação já mencionada:
\begin{center}
\begin{picture}(155.88,120)(0,0)
\DashArrowLine(0,0)(51.96,30){5}
\Text(15.98,15)[r]{$P$}
\DashArrowLine(51.96,30)(51.96,90){5}
\Text(41.96,60)[r]{$P-k$}
\DashArrowLine(51.96,90)(0,120){5}
\Text(15.98,105)[r]{$P'=P-q$}
\Gluon(51.96,30)(103.92,60){3}{6}
\Text(77.94,25)[c]{$k$}
\LongArrow(73.61,52.5)(82.27,57.5)
\Text(51.96,10)[l]{$\lambda\rho$}
\Text(103.92,40)[r]{$\alpha\beta$}
\Gluon(103.92,60)(51.96,90){-3}{6}
\Text(77.94,95)[c]{$k-q$}
\LongArrow(82.27,62.5)(73.61,67.5)
\Text(51.96,110)[l]{$\sigma\tau$}
\Text(103.92,80)[r]{$\mu\nu$}
\DashArrowLine(155.88,30)(103.92,60){5}
\Text(139.9,45)[l]{$p$}
\DashArrowLine(103.92,60)(155.88,90){5}
\Text(139.9,75)[l]{$p'=p+q$}
\end{picture}
\end{center}
\begin{eqnarray*}
i\mathcal{M}^{(2)}&=&i\int\frac{d^4k}{(2\pi)^4}V^{\sigma\tau}(P',P-k)\frac{i\mathcal{P}_{\sigma\tau,\mu\nu}}{(k-q)^2}\frac{i}{(P-k)^2-M^2}\\&&\times\frac{i\mathcal{P}_{\lambda\rho,\alpha\beta}}{k^2}V^{\lambda\rho}(P-k,P)V^{\mu\nu,\alpha\beta}(p',p)\\&\approx&-\frac{\kappa^2M^4}{4}\underbrace{(\eta_{\mu\nu}-2\delta^0_\mu\delta^0_\nu)(\eta_{\alpha\beta}-2\delta^0_\alpha\delta^0_\beta)V^{\mu\nu,\alpha\beta}(p',p)}_{i\kappa^2\left(4E^2-\frac{{\bf q}^2}{2}\right)}\\&&\times\underbrace{\int\frac{d^4k}{(2\pi)^4}\frac{1}{k^2(k-q)^2[(P-k)^2-M^2]}}_{J\approx\frac{-i}{32M|{\bf q}|}}\\&\approx&\frac{-8\pi^2 G^2M^3}{|{\bf q}|}\left(4E^2-\frac{{\bf q}^2}{2}\right)
\end{eqnarray*}
onde usamos as seguintes aproximações
\begin{eqnarray*}
i\mathcal{P}_{\sigma\tau,\mu\nu}V^{\sigma\tau}(P',P-k)&=&\frac{\kappa}{2}[P'_\mu (P-k)_\nu+P'_\nu (P-k)_\mu-\eta_{\mu\nu}M^2]\\&\approx&-\frac{\kappa M^2}{2}(\eta_{\mu\nu}-2\delta^0_\mu\delta^0_\nu)\\i\mathcal{P}_{\lambda\rho,\alpha\beta}V^{\lambda\rho}(P-k,P)&=&\frac{\kappa}{2}[(P-k)_\alpha P_\beta+(P-k)_\beta P_\alpha-\eta_{\alpha\beta}M^2]\\&\approx&-\frac{\kappa M^2}{2}(\eta_{\alpha\beta}-2\delta^0_\alpha\delta^0_\beta)
\end{eqnarray*}
que são válidas porque termos como $PPPP$, $kPPP$ e $kkPP$ no numerador da integral sobre $k$ contribuem com fatores $M^3$, $M^2$ e $M$ respectivamente (veja Apêndice B).
\begin{center}
\begin{picture}(215.88,120)(0,0)
\DashArrowLine(0,0)(51.96,30){5}
\Text(15.98,15)[r]{$P$}
\DashArrowLine(51.96,30)(51.96,90){5}
\Text(41.96,60)[r]{$P-k$}
\DashArrowLine(51.96,90)(0,120){5}
\Text(15.98,105)[r]{$P'=P-q$}
\Gluon(51.96,30)(103.92,60){3}{6}
\Text(77.94,25)[c]{$k$}
\LongArrow(73.61,52.5)(82.27,57.5)
\Text(51.96,10)[l]{$\gamma\delta$}
\Text(103.92,40)[r]{$\alpha\beta$}
\Gluon(103.92,60)(51.96,90){-3}{6}
\Text(77.94,95)[c]{$k-q$}
\LongArrow(82.27,62.5)(73.61,67.5)
\Text(51.96,110)[l]{$\epsilon\zeta$}
\Text(103.92,80)[r]{$\lambda\rho$}
\Gluon(103.92,60)(163.92,60){-3}{6}
\Text(133.92,50)[c]{$q$}
\LongArrow(128.92,70)(138.92,70)
\Text(103.92,50)[l]{$\mu\nu$}
\Text(163.92,50)[r]{$\sigma\tau$}
\DashArrowLine(215.88,30)(163.92,60){5}
\Text(199.9,45)[l]{$p$}
\DashArrowLine(163.92,60)(215.88,90){5}
\Text(199.9,75)[l]{$p'=p+q$}
\end{picture}
\end{center}
\begin{eqnarray*}
i\mathcal{M}^{(3)}&=&i\int\frac{d^4k}{(2\pi)^4}V^{\epsilon\zeta}(P',P-k)\frac{i\mathcal{P}_{\epsilon\zeta,\lambda\rho}}{(k-q)^2}\frac{i}{(P-k)^2-M^2}\frac{i\mathcal{P}_{\gamma\delta,\alpha\beta}}{k^2}V^{\gamma\delta}(P-k,P)\\&&\times V^{\alpha\beta,\mu\nu,\lambda\rho}(k,q,k-q)\frac{i\mathcal{P}_{\mu\nu,\sigma\tau}}{q^2}V^{\sigma\tau}(p',p)\\&\approx&\frac{\kappa^3M^4}{8{\bf q}^2}(p'_\mu p_\nu+p'_\nu p_\mu-\eta_{\mu\nu}m^2)\\&&\times\int\frac{d^4k}{(2\pi)^4}\frac{(\eta_{\alpha\beta}-2\delta^0_\alpha\delta^0_\beta)V^{\alpha\beta,\mu\nu,\lambda\rho}(k,q,k-q)(\eta_{\lambda\rho}-2\delta^0_\lambda\delta^0_\rho)}{k^2(k-q)^2[(P-k)^2-M^2]}\\&=&\frac{\kappa^3M^4}{8{\bf q}^2}(p'_\mu p_\nu+p'_\nu p_\mu-\eta_{\mu\nu}m^2)\int\frac{d^4k}{(2\pi)^4}\frac{i\kappa}{k^2(k-q)^2[(P-k)^2-M^2]}\\&&\times[2(2k\cdot q-q^2-2k^2)\delta^\mu_0\delta^\nu_0+2q^\mu q^\nu-2k^\mu k^\nu+q^\mu k^\nu+q^\nu k^\mu\\&&+(k^2-k\cdot q)\eta^{\mu\nu}]\\&\approx&\frac{\kappa^3M^4}{8{\bf q}^2}(p'_\mu p_\nu+p'_\nu p_\mu-\eta_{\mu\nu}m^2)\frac{\kappa}{128M|{\bf q}|}(9q^\mu q^\nu-q^2\eta^{\mu\nu}-q^2\delta^\mu_0\delta^\nu_0)\\&=&\frac{-8\pi^2 G^2M^3}{|{\bf q}|}\left(-m^2-\frac{{\bf p}^2}{4}+\frac{7{\bf q}^2}{16}\right)
\end{eqnarray*}
\begin{center}
\begin{picture}(163.92,120)(0,0)
\DashArrowLine(0,0)(51.96,30){5}
\Text(15.98,15)[r]{$P$}
\DashArrowLine(51.96,30)(51.96,90){5}
\Text(41.96,60)[r]{$P-k$}
\DashArrowLine(51.96,90)(0,120){5}
\Text(15.98,105)[r]{$P'=P-q$}
\Gluon(51.96,30)(111.96,30){3}{6}
\LongArrow(76.96,20)(86.96,20)
\Text(81.96,40)[c]{$k$}
\Text(51.96,20)[l]{$\lambda\rho$}
\Text(111.96,20)[r]{$\alpha\beta$}
\Gluon(111.96,90)(51.96,90){3}{6}
\LongArrow(86.96,100)(76.96,100)
\Text(81.96,80)[c]{$k-q$}
\Text(51.96,100)[l]{$\sigma\tau$}
\Text(111.96,99)[r]{$\mu\nu$}
\DashArrowLine(163.92,0)(111.96,30){5}
\Text(147.94,15)[l]{$p$}
\DashArrowLine(111.96,30)(111.96,90){5}
\Text(121.96,60)[l]{$p+k$}
\DashArrowLine(111.96,90)(163.92,120){5}
\Text(147.94,105)[l]{$p'=p+q$}
\end{picture}
\end{center}
\begin{center}
\begin{picture}(163.92,120)(0,0)
\DashArrowLine(0,0)(51.96,30){5}
\Text(15.98,15)[r]{$P$}
\DashArrowLine(51.96,30)(51.96,90){5}
\Text(41.96,60)[r]{$P-k$}
\DashArrowLine(51.96,90)(0,120){5}
\Text(15.98,105)[r]{$P'=P-q$}
\Gluon(51.96,30)(111.96,90){3}{6}
\LongArrow(53.42,41.46)(60.5,48.54)
\Text(66.96,35)[l]{$k$}
\Text(51.96,20)[l]{$\lambda\rho$}
\Text(111.96,20)[r]{$\alpha\beta$}
\Gluon(111.96,30)(51.96,90){-3}{6}
\LongArrow(60.5,71.46)(53.42,78.54)
\Text(66.96,85)[l]{$k-q$}
\Text(51.96,100)[l]{$\sigma\tau$}
\Text(111.96,99)[r]{$\mu\nu$}
\DashArrowLine(163.92,0)(111.96,30){5}
\Text(147.94,15)[l]{$p$}
\DashArrowLine(111.96,30)(111.96,90){5}
\Text(121.96,60)[l]{$p'-k$}
\DashArrowLine(111.96,90)(163.92,120){5}
\Text(147.94,105)[l]{$p'=p+q$}
\end{picture}
\end{center}
\begin{eqnarray*}
i\mathcal{M}^{(4)}&=&i\int\frac{d^4k}{(2\pi)^4}V^{\sigma\tau}(P',P-k)\frac{i\mathcal{P}_{\sigma\tau,\mu\nu}}{(k-q)^2-\mu^2}\frac{i}{(P-k)^2-M^2}\frac{i\mathcal{P}_{\lambda\rho,\alpha\beta}}{k^2-\mu^2}\\&&\times V^{\lambda\rho}(P-k,P)V^{\mu\nu}(p',p+k)\frac{i}{(p+k)^2-m^2}V^{\alpha\beta}(p+k,p)\\&&+i\int\frac{d^4k}{(2\pi)^4}V^{\sigma\tau}(P',P-k)\frac{i\mathcal{P}_{\sigma\tau,\alpha\beta}}{(k-q)^2-\mu^2}\frac{i}{(P-k)^2-M^2}\frac{i\mathcal{P}_{\lambda\rho,\mu\nu}}{k^2-\mu^2}\\&&\times V^{\lambda\rho}(P-k,P)V^{\mu\nu}(p',p'-k)\frac{i}{(p'-k)^2-m^2}V^{\alpha\beta}(p'-k,p)\\
&\approx&-i\frac{\kappa^2M^4}{4}\times\\
&&\left\{\int\frac{d^4k}{(2\pi)^4}\frac{(\eta_{\mu\nu}-2\delta^0_\mu\delta^0_\nu)V^{\mu\nu}(p',p+k)(\eta_{\alpha\beta}-2\delta^0_\alpha\delta^0_\beta)V^{\alpha\beta}(p+k,p)}{(k^2-\mu^2)[(k-q)^2-\mu^2][(P-k)^2-M^2][(p+k)^2-m^2]}\right.\\
&&\left.+\int\frac{d^4k}{(2\pi)^4}\frac{(\eta_{\mu\nu}-2\delta^0_\mu\delta^0_\nu)V^{\mu\nu}(p',p'-k)(\eta_{\alpha\beta}-2\delta^0_\alpha\delta^0_\beta)V^{\alpha\beta}(p'-k,p)}{(k^2-\mu^2)[(k-q)^2-\mu^2][(P-k)^2-M^2][(p'-k)^2-m^2]}\right\}\\
&\approx&i64\pi^2G^2M^4(2m^2+4{\bf p}^2)^2\times\\
&&\left\{\int\frac{d^4k}{(2\pi)^4}\frac{1}{(k^2-\mu^2)[(k-q)^2-\mu^2][(P-k)^2-M^2][(p+k)^2-m^2]}\right.\\
&&\left.+\int\frac{d^4k}{(2\pi)^4}\frac{1}{(k^2-\mu^2)[(k-q)^2-\mu^2][(P-k)^2-M^2][(p'-k)^2-m^2]}\right\}\\
&\approx&i64\pi^2G^2M^4(2m^2+4{\bf p}^2)^2\underbrace{(K+K')}_{\approx\frac{-1}{8\pi M{\bf q}^2|{\bf p}|}\ln\frac{|{\bf q}|}{\mu}}\\
&\approx&-i\frac{8\pi G^2M^3}{|{\bf p}|{\bf q}^2}(2m^2+4{\bf p}^2)^2\ln \frac{|{\bf q}|}{\mu}
\end{eqnarray*}
onde usamos
\begin{eqnarray*}
(\eta_{\mu\nu}-2\delta^0_\mu\delta^0_\nu)V^{\mu\nu}(p',p+k)=(\eta_{\alpha\beta}-2\delta^0_\alpha\delta^0_\beta)V^{\alpha\beta}(p+k,p)&\approx&\frac{i\kappa}{2}(2m^2+4{\bf p}^2)\\
(\eta_{\mu\nu}-2\delta^0_\mu\delta^0_\nu)V^{\mu\nu}(p',p'-k)=(\eta_{\alpha\beta}-2\delta^0_\alpha\delta^0_\beta)V^{\alpha\beta}(p'-k,p)&\approx&\frac{i\kappa}{2}(2m^2+4{\bf p}^2).
\end{eqnarray*}
Assim conseguimos mostrar que até o nível de loop
\[\mathcal{M}_{\mbox{semiclássica}}\approx\frac{\mathcal{M}_{\mbox{efetiva}}}{2M}\]
ou seja, as duas teorias predizem o mesmo resultado no limite em que uma das massas envolvidas é muito grande \cite{Accioly4}.
\section{Potencial em Ordem $G^2$}
Usando a definição do potencial não-relativístico \cite{Iwasaki,Gupta80}
\begin{eqnarray*}
V(r)&=&\left.\int\frac{d^3q}{(2\pi)^3}e^{i{\bf q}\cdot{\bf r}}\left[\frac{i\mathcal{M}^{total}}{4M\sqrt{E'E}}-\int\frac{d^3p''}{(2\pi)^3}\frac{\frac{i\mathcal{M}^{(1)}({\bf p}',{\bf p}'')}{4M\sqrt{E'E''}}\frac{i\mathcal{M}^{(1)}({\bf p}'',{\bf p})}{4M\sqrt{E''E}}}{E-E''}\right]\right|_{E'=E}\\
&=&\int\frac{d^3q}{(2\pi)^3}e^{i{\bf q}\cdot{\bf r}}\left[-\frac{4\pi GMm}{{\bf q}^2}\left(1+\frac{3{\bf p}^2}{2m^2}\right)-i\frac{8\pi G^2M^2m^3}{|{\bf p}|{\bf q}^2}\ln\frac{|{\bf q}|}{\mu}\right.\\&&\left.-\frac{6\pi^2G^2M^2m}{|{\bf q}|}-\left(-i\frac{8\pi G^2M^2m^3}{|{\bf p}|{\bf q}^2}\ln\frac{|{\bf q}|}{\mu}-\frac{7\pi^2G^2M^2m}{|{\bf q}|}\right)\right]
\end{eqnarray*}
onde o último termo é a subtração da segunda aproximação de Born. O potencial resultante é nada mais que o potencial não-relativístico em coordenadas harmônicas (ou seja, no gauge de de Donder empregado)
\[V(r)=-\frac{GMm}{r}-\frac{3GM{\bf p}^2}{2mr}+\frac{G^2M^2m}{2r^2}.\]

\chapter{Aplicações}
\section{Massa do fóton}
Uma das primeiras aplicações que podemos fazer é obtida já em primeira ordem de perturbação: trata-se do cálculo da massa do fóton.

Igualando a seção de choque do fóton massivo (Proca) obtida no Capítulo~2 para ângulos pequenos (apesar de que qualquer um dos resultados para partículas massivas serviria igualmente, pois são idênticos neste limite, veja Tabela 2.1) e massas pequenas $\frac{m}{E}\ll 1$
\[\frac{d\sigma}{d\Omega}\approx\left(\frac{4GM}{\theta^2}\right)^2\left(1+\frac{m^2}{E^2}\right)\]
e a bem conhecida aproximação clássica para a seção de choque \cite{Landau}
\[\frac{d\sigma}{d\Omega}\approx-\frac{b}{\theta}\frac{db}{d\theta}\]
onde $b$ é o parâmetro de impacto, obtemos um ângulo de desvio para partículas massivas
\[\theta=\frac{4GM}{b}\left(1+\frac{m^2}{2E^2}\right)\]
que, comparado com o desvio clássico das partículas (rasantes\footnote{Lembrando que $\theta=\frac{2GM}{b}(1+\cos\phi)$ é o ângulo clássico de desvio para partículas incidentes em qualquer posição em relação a M \cite{MTW}, e que se reduz a $\theta=\frac{4GM}{b}$ apenas no caso em que as partículas passam rasantes a M, ou seja, $\phi\rightarrow 0$. O ângulo $\phi$ é conhecido como {\it elongação}, veja Figura~5.2.}) sem massa $\theta_c=\frac{4GM}{b}$ nos dá um limite para a massa do fóton \cite{Paszko}
\[m=2\pi\nu\sqrt{\frac{2\Delta\theta}{\theta_c}}\]
onde $\Delta\theta=\theta-\theta_c$.

Esta expressão pode ser obtida classicamente (veja Apêndice C e referências \cite{Wheelon,MTW}, ou ainda referência \cite{Lowenthal} para uma derivação heurística).

Usando os dados do desvio da luz \cite{Shapiro} para uma freqüência de $\nu=2GHz$ e igualando $\Delta\theta/\theta_c$ com o erro relativo $1.4\times 10^{-4}$, obtemos uma massa $m=1.4\times 10^{-7}eV=2.5\times 10^{-41}g$, um valor muito mais conservador do que o obtido por outros métodos para se estimar a massa do fóton \cite{Luo,PDG}.
\section{Violação do Princípio da Equivalência}
Se adotarmos agora a massa do fóton como exatamente nula, vamos encontrar modificações em 
relação ao caso clássico somente em segunda ordem de perturbação.

No Capítulo 3, concluímos que as seções de choque em segunda ordem possuem uma dependência 
explícita na energia da partícula. Em particular, estas expressões (tanto para o caso da partícula escalar sem massa quanto para o fóton) se reduzem para ângulos pequenos a
\[\frac{d\sigma}{d\Omega}\approx\left(\frac{4GM}{\theta^2}\right)^2\left(1+\frac{15\pi GME\theta}{8}\right).\]

Usando novamente a aproximação clássica da seção de choque, obtemos o seguinte ângulo de desvio
\[\theta_{Efetiva}=\frac{4GM}{b}+\frac{60\pi^2G^3M^3}{\lambda b^2}\]
válido\footnote{Entendemos o argumento apontado por Drummond e Hathrell na Referência \cite{Drummond}, sobre a validade deste método e por isso afirmamos que este termo adicional representa apenas uma minúscula perturbação ao resultado clássico, portanto, não acarretando termos do tipo $\ln\theta$ que obviamente não são corretos.} para $\lambda\gg GM$, e assim temos
\[\frac{\Delta\theta}{\theta_c}=\frac{15\pi^2G^2M^2}{\lambda b}.\]

Por outro lado, podemos também calcular o desvio da luz no caso em que $\lambda\ll b$, usando a Teoria Semiclássica através do método JWKB (veja Apêndice~D e referências \cite{Sanchez,Accioly5}) obtendo
\[\theta_{JWKB}=\frac{4GM}{b}+\frac{\lambda^2}{32\pi b^2}\]
e portanto
\[\frac{\Delta\theta}{\theta_c}=\frac{\lambda^2}{128\pi GMb}.\]

Na Figura 5.1 temos $\Delta\theta/\theta_c$ em função de $\lambda$ para o caso de partículas rasantes ao Sol, ou seja, $GM_\odot=1.476625\times 10^3m$ e $b=R_\odot=6.961\times 10^8m$.
\begin{figure}[!h]
\begin{center}
\centering
\includegraphics[angle=-90,scale=.59]{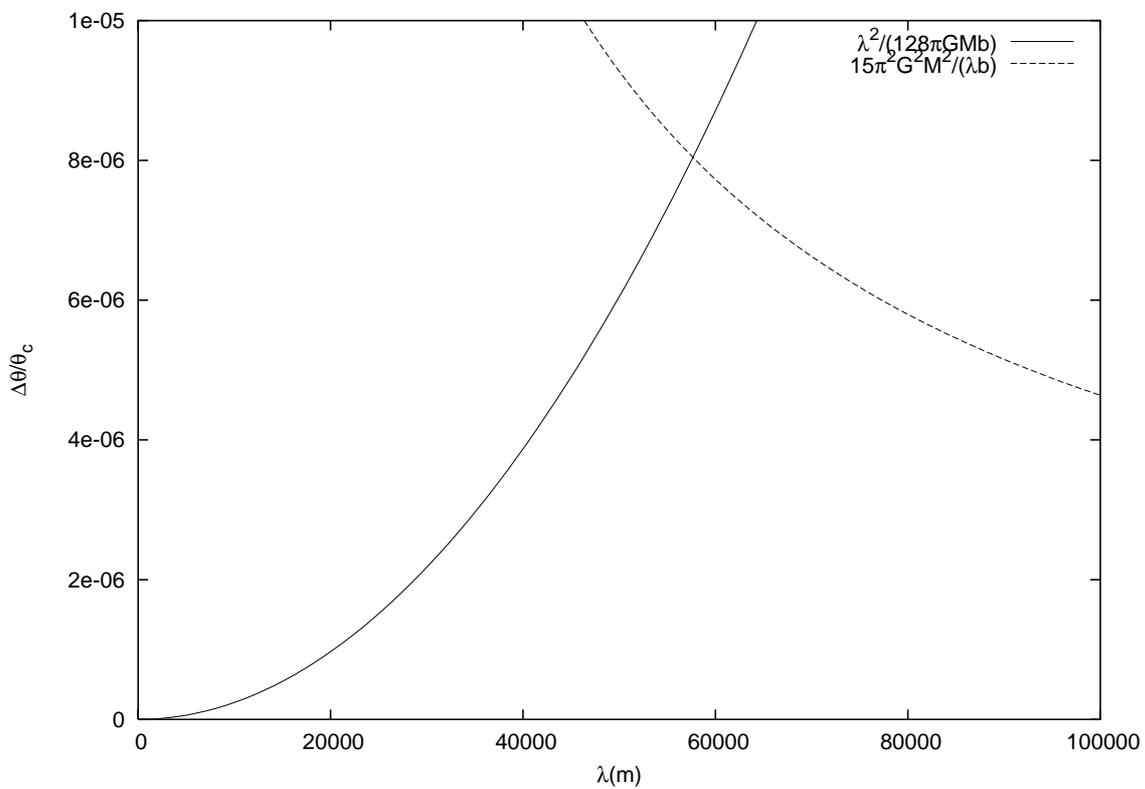}
\end{center}
\caption{\small $\Delta\theta/\theta_c$ em função de $\lambda$ (em metros).}
\end{figure}

Algumas observações podem ser feitas aqui:
\begin{itemize}
\item[-] Existe uma dependência explícita no comprimento de onda $\lambda$ (ou na energia) da partícula, o que demonstra uma violação do Princípio da Equivalência;
\item[-] Os efeitos são mais intensos a medida que $\lambda$ se aproxima de $4\pi\sqrt[3]{30}GM$ (no caso do Sol $\lambda=5.7657\times 10^4m$), contrariando o senso comum de que os efeitos quânticos gravitacionais surgem apenas em altíssimas energias tais como a energia de Planck ($E_{Planck}\sim 10^{19}GeV$ ou $\lambda_{Planck}\sim 10^{-35}m$);
\item[-] Infelizmente para comprimentos de onda típicos, os efeitos são minúsculos, na verdade impossíveis de serem detectados com a tecnologia atual (de fato, $\Delta\theta/\theta_c\sim 10^{-17}$ para $\lambda\sim 0.1m$, enquanto que a melhor precisão experimental atual \cite{Shapiro} é $\Delta\theta/\theta_c\sim 10^{-4}$).
\end{itemize}
\section{Lentes Gravitacionais}
Nossa última aplicação diz respeito às lentes gravitacionais. Na Figura 5.2 temos um esboço da geometria envolvida.
\begin{figure}[!h]
\begin{center}
\includegraphics[scale=.8]{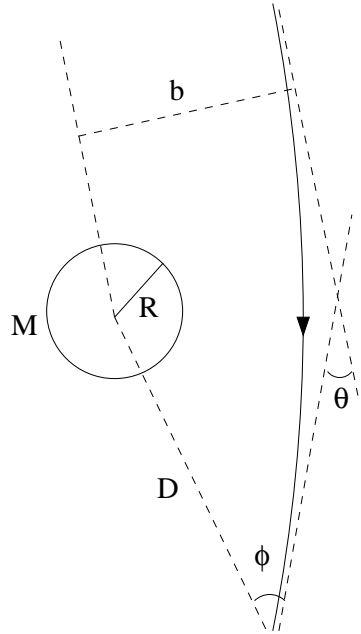}
\end{center}
\caption{\small Geometria da deflexão da luz pelo Sol. A elongação $\phi$ é o ângulo entre o Sol e a imagem da estrela vista da Terra.}
\end{figure}
\subsection{Anel de Einstein}
O tamanho (angular) clássico do Anel de Einstein pode ser facilmente obtido através da condição de alinhamento total \cite{Ohanian}
\begin{eqnarray*}
\theta_{Anel}&=&\phi\\
\frac{4GM}{b}&\approx &\frac{b}{D}
\end{eqnarray*}
portanto $b=\sqrt{4GMD}$, e assim temos
\[\theta_{Anel}=\sqrt{\frac{4GM}{D}}.\]

De maneira análoga obtemos para a Teoria Semiclássica (JWKB)
\[\theta_{JWKB}=\sqrt{\frac{4GM}{D}}+\frac{\lambda^2}{256\pi GMD}\]
e assim
\[\frac{\Delta\theta}{\theta_{Anel}}=\frac{\lambda^2}{512\pi\sqrt{G^3M^3D}}\]
válido para $\lambda$ pequeno. Vemos então que as lentes gravitacionais são {\it cromáticas}, porém o efeito é minúsculo, pois para uma galáxia típica $M\sim 10^{11}M_\odot$ e $D\sim 10^{10}ly$ (e portanto $\theta_{Anel}\sim 1\;arcsec$), e para uma freqüência típica $\nu\sim 1GHz$ temos $\Delta\theta/\theta_{Anel}\sim 10^{-38}$ que é impossível de ser observado.

\chapter{Epílogo}
É quase um consenso entre os físicos que a Mecânica Quântica e a Relatividade Geral são os dois grandes pilares sobre os quais se assenta a Física Moderna. No entanto, apesar dos hercúleos esforços despendidos por um número considerável de físicos ilustres, não se conseguiu até agora a fusão destas duas teorias. Qual seria, ao menos em princípio, a causa principal desta incompatibilidade? Talvez a dificuldade conceitual de reconciliar uma teoria local, como a Relatividade Geral, com efeitos quânticos não locais ou, equivalentemente, de reconciliar o caráter local do Princípio da Equivalência Forte --- {\it Um observador ideal imerso em um campo gravitacional pode escolher um sistema de referência no qual a gravitação passa desapercebida} \cite{Aldrovandi} --- com o caráter não local do Princípio da Incerteza \cite{Lammerzahl}. É digno de nota que este fato fundamental tem sido desprezado por um grande número de investigadores em sua cruzada pela Gravitação Quântica. Por outro lado, os dois únicos caminhos de acesso à Gravitação Quântica que fornecem uma descrição matemática completa das propriedades quânticas do campo gravitacional, ou seja, {\it String Theory} e {\it Loop Quantum Gravity}, não foram ainda testados. Isto não é de causar espécie já que a escala de energia característica da Gravitação Quântica é a energia de Planck, $E_{\mathrm P} =10^{19} GeV$, um valor que está tão longe do alcance experimental que a idéia de testes observacionais diretos parece, há muito, não passar de um plano quimérico. Certamente as reivindicações constantes da literatura sobre o fato de que nos encontramos no alvorecer da fenomenologia da Gravitação Quântica (veja, por exemplo, Amelino-Camelia \cite{Amelino}), são ainda bastante especulativas, o que parece de certa maneira ameaçar o futuro deste enfoque fenomenológico.

Neste trabalho, nos devotamos à análise da incompatibilidade existente entre o Princípio da Equivalência Fraco (universalidade da queda livre, ou igualdade das massas inercial e gravitacional) --- considerado a pedra angular tanto da Gravitação Newtoniana quanto da Relatividade Geral --- e a Mecânica Quântica. É importante frisar, contudo, que a equivalência dos princípios da universalidade da queda livre ({\it Princípio da Equivalência de Galileu}) e da igualdade das massas inercial e gravitacional ({\it Princípio da Equivalência de Newton}), suposta válida no contexto das teorias gravitacionais acima mencionadas, não é suportada pela experiência. De fato, os experimentos em nível clássico que dão robustez ao último somente dão suporte limitado ao primeiro \cite{Ohanian}. É conveniente, pois, que eles sejam tratados como princípios distintos. No domínio quântico, todavia, as coisas são bastante diferentes: teoria e experiência concordam no que concerne à Equivalência de Newton, mas ambas estão em desacordo com a Equivalência de Galileu. Realmente, usando um interferômetro de nêutrons, Colella, Overhauser e Werner \cite{COW} observaram que o deslocamento quanto-mecânico causado pela interação do nêutron com o campo gravitacional da Terra (suposto ser Newtoniano) era dependente da massa do nêutron. Esta notável experiência, conhecida por razões óbvias como experiência COW, confirma a idéia de que $m_{\mathrm I} = m_{\mathrm G}$ (Equivalência de Newton). Por outro lado ela mostra que existe um conflito inevitável entre a Equivalência de Galileu e a Mecânica Quântica quando a gravitação é descrita pela teoria de Newton. É surpreendente que a existência de efeitos de interferência dependentes da massa tenham sido previstos por Greenberger \cite{Greenberger} uns poucos anos antes da experiência COW ao aplicar a Mecânica Quântica ao problema de uma partícula ligada a um potencial gravitacional externo. Vale a pena mencionar que a igualdade entre as massas gravitacional e inercial estava incorporada em seu artigo seminal. Ele também supôs que o campo externo que atuava sobre a partícula era descrito pela gravitação Newtoniana. Em 1986, no entanto, uma reanálise dos dados de Eötvös feita por Fischbach et al. \cite{Fischbach2} suscitou pela primeira vez dúvidas sobre a validade da Equivalência de Newton. Os dados pareciam indicar a existência de uma nova força --- a quinta força. Apesar dos prodigiosos esforços despendidos pelos experimentais, nada foi encontrado porém que pudesse confirmar que esta força era realmente verossímil \cite{Nelson}. Conseqüentemente, no que concerne à campos gravitacionais externos descritos pela gravitação de Newton, somente a Equivalência de Newton desfruta do status de um principio básico tanto no domínio clássico quanto no quântico.

Ao longo deste trabalho admitimos sempre a validade do Princípio da Equivalência de Newton, e trabalhamos com a versão linearizada da gravitação de Einstein, em vez da gravitação de Newton. Encontramos, mesmo assim, já em primeira ordem de perturbação, que as seções de choque são distintas para partículas de spins diferentes. Quando os cálculos são levados até a próxima ordem de perturbação, as seções de choque passam a depender explicitamente da energia, o que acarreta, conseqüentemente, que tanto a deflexão gravitacional quanto o tamanho angular do anel de Einstein herdem esta dependência na energia. Certamente, isto também se aplica à Teoria Efetiva da Gravitação na aproximação em que uma das massas é muito grande, pois, conforme mostramos, nesta aproximação ela é equivalente à Teoria Semiclássica. Isto nos permite afirmar com segurança que as teorias Semiclássica e Efetiva da Gravitação são incompatíveis com o Princípio da Equivalência de Galileu e, portanto, com o Princípio da Equivalência Fraco. Estas violações da Equivalência Fraca que encontramos são, porém, muito pequenas. No caso da deflexão gravitacional de fótons pelo Sol, os efeitos quânticos são da ordem de $10^{-17}\; arcsec$ , o que impede que eles sejam detectados com a tecnologia atual. Espera-se que, em meados de 2010, a precisão experimental concernente a medidas astronômicas atinja a casa dos $nano-arcsec$ \cite{NASA}, permitindo provavelmente a observação destes efeitos.

O que teria a teoria de cordas para nos dizer sobre a universalidade da queda livre? As cordas são objetos estendidos e como tal estão sujeitas a forças de maré, e portanto não obedecem a este princípio.

Conclui-se da exposição acima que o Princípio da Equivalência --- seja em sua forma forte ou fraca --- é um dos fatores responsáveis pela existência de um conflito central entre a Relatividade Geral e a Mecânica Quântica. Uma saída para este dilema seria talvez considerar a gravitação sem o Princípio da Equivalência \cite{Aldrovandi2}. À primeira vista esta parece ser uma atitude muito drástica no que concerne à Relatividade Geral. Felizmente, a versão Teleparalela da Relatividade Geral --- uma descrição da interação gravitacional semelhante à força de Lorentz do eletromagnetismo e, que em decorrência nada tem a ver com o Princípio da Equivalência --- é equivalente à Relatividade Geral, e fornece em conseqüência uma teoria consistente de gravitação na ausência deste princípio \cite{Hammond}.

\appendix
\chapter{Integrais em $3D$}
\section{Cálculo de $I$, $I^i$ e $I^{ij}$}
Vamos precisar da seguinte integral básica que pode ser calculada pelo método de resíduos \cite{Dalitz}
\[\lim_{\epsilon\rightarrow 0}\int\frac{d^3k}{[({\bf k}-{\bf P})^2+\Lambda^2]({\bf p}^2-{\bf k}^2+i\epsilon)}=\frac{i\pi^2}{P}\ln\frac{p-P+i\Lambda}{p+P+i\Lambda}\]
derivando em relação à $\Lambda$ e a $P^i$ obtemos respectivamente (fica subentendido o limite $\epsilon\rightarrow 0$)
\begin{eqnarray*}
\int\frac{d^3k}{[({\bf k}-{\bf P})^2+\Lambda^2]^2({\bf p}^2-{\bf k}^2+i\epsilon)}&=&\frac{\pi^2}{\Lambda[(p+i\Lambda)^2-P^2]}\\
\int\frac{k^id^3k}{[({\bf k}-{\bf P})^2+\Lambda^2]^2({\bf p}^2-{\bf k}^2+i\epsilon)}&=&P^i\left\{\frac{\pi^2}{\Lambda[(p+i\Lambda)^2-P^2]}-\frac{i\pi^2}{2P^3}\ln\frac{p-P+i\Lambda}{p+P+i\Lambda}\right.\\&&\left.-\frac{i\pi^2(p+i\Lambda)}{P^2[(p+i\Lambda)^2-P^2]}\right\}
\end{eqnarray*}
e ainda integrando a última equação por $\Lambda^2$ obtemos
\[\int\frac{k^id^3k}{[({\bf k}-{\bf P})^2+\Lambda^2]({\bf p}^2-{\bf k}^2+i\epsilon)}=\frac{\pi^2P^i}{P}\left(\frac{\Lambda+ip}{P}+i\frac{p^2+P^2+\Lambda^2}{2P^2}\ln\frac{p-P+i\Lambda}{p+P+i\Lambda}\right)\]
e derivando em relação a $P^j$ temos
\newpage
\begin{eqnarray*}
\int\frac{k^ik^jd^3k}{[({\bf k}-{\bf P})^2+\Lambda^2]^2({\bf p}^2-{\bf k}^2+i\epsilon)}&=&\frac{\pi^2\delta^{ij}}{2P}\left(i\frac{p^2+P^2+\Lambda^2}{2P^2}\ln\frac{p-P+i\Lambda}{p+P+i\Lambda}\right.\\&&\left.+\frac{\Lambda+ip}{P}\right)-\frac{\pi^2P^iP^j}{2P}\left[\frac{2(\Lambda+ip)}{P^3}\right.\\&&\left.+\frac{3i}{2}\left(\frac{p^2+P^2+\Lambda^2}{P^4}\right)\ln\frac{p-P+i\Lambda}{p+P+i\Lambda}\right.\\&&\left.+\frac{i(p^2+3P^2+\Lambda^2)(p+i\Lambda)}{P^3[(p+i\Lambda)^2-P^2]}\right.\\&&\left.-\frac{2P}{\Lambda[(p+i\Lambda)^2-P^2]}\right].
\end{eqnarray*}

As integrais necessárias neste trabalho são da forma
\[(I,I^i,I^{ij})=\int\frac{(1,k^i,k^ik^j)d^3k}{[({\bf p'}-{\bf k})^2+\mu^2][({\bf p}-{\bf k})^2+\mu^2]({\bf p}^2-{\bf k}^2+i\epsilon)}\]
que podem ser obtidas a partir das integrais anteriores através da identidade de Feynman
\[\frac{1}{ab}=\int_{-1}^1\frac{dz}{2}\left[\frac{a(1+z)+b(1-z)}{2}\right]^{-2}\]
agrupando-se os denominadores de ${\bf p}$ e ${\bf p'}$ tal que ${\bf P}=\frac{{\bf p+p'}}{2}+z\frac{{\bf p-p'}}{2}$, $P^2=p^2(\cos^2\frac{\theta}{2}+z^2\sin^2\frac{\theta}{2})$ e $\Lambda^2=p^2(1-z^2)\sin^2\frac{\theta}{2}+\mu^2$ onde usamos ${\bf p}^2={\bf p'}^2=p^2$ e ${\bf p}\cdot{\bf p'}=p^2\cos\theta$.
As integrais finais são puramente algébricas, porém tediosas. O importante para nós é simplesmente o valor destas expressões no limite em que $\mu\rightarrow 0$, portanto
\[I\approx-\frac{i\pi^2}{2p^3\sin^2\frac{\theta}{2}}\ln\frac{2p\sin\frac{\theta}{2}}{\mu}\]
$I^i=\frac{({\bf p'+p})^i}{2}J$ onde \[J\approx\frac{\pi^3}{4p^3\cos^2\frac{\theta}{2}}\left(1-\frac{1}{\sin\frac{\theta}{2}}\right)-\frac{i\pi^2}{2p^3\cos^2\frac{\theta}{2}}\left(\ln\frac{\mu}{2p}+\frac{1}{\sin^2\frac{\theta}{2}}\ln\frac{2p\sin\frac{\theta}{2}}{\mu}\right)\]
\newpage
\noindent
$I^{ij}=A\delta^{ij}+B{\bf q}^i{\bf q}^j+C({\bf p'+p})^i({\bf p'+p})^j$ onde
\begin{eqnarray*}
A&\approx&\frac{\pi^2}{4p}\left(\frac{-\pi}{1+\sin\frac{\theta}{2}}+\frac{2i\ln\sin\frac{\theta}{2}}{\cos^2\frac{\theta}{2}}\right)\\
B&\approx&\frac{-\pi^2}{16p^3\sin\frac{\theta}{2}\cos^2\frac{\theta}{2}}\left[\pi\left(1-\sin\frac{\theta}{2}\right)+2i\sin\frac{\theta}{2}\ln\sin\frac{\theta}{2}\right]-\frac{i\pi^2}{8p^3\sin^2\frac{\theta}{2}}\ln\frac{2p\sin\frac{\theta}{2}}{\mu}\\
C&\approx&\frac{-\pi^2}{16p^3\sin\frac{\theta}{2}\cos^4\frac{\theta}{2}}\left[2i\sin\frac{\theta}{2}\cos^2\frac{\theta}{2}+\pi\left(2-3\sin\frac{\theta}{2}+\sin^3\frac{\theta}{2}\right)\right.\\&&\left.-2i\sin\frac{\theta}{2}\left(\sin^2\frac{\theta}{2}-3\right)\ln\sin\frac{\theta}{2}\right]-\frac{i\pi^2}{8p^3\sin^2\frac{\theta}{2}}\ln\frac{2p\sin\frac{\theta}{2}}{\mu}
\end{eqnarray*}
\section{Integrais Úteis}
\begin{eqnarray*}
\int\frac{d^3q}{(2\pi)^3}\frac{e^{i{\bf q}\cdot{\bf r}}}{|{\bf q}|}&=&\frac{1}{2\pi^2r^2}\\
\int\frac{d^3q}{(2\pi)^3}\frac{e^{i{\bf q}\cdot{\bf r}}}{{\bf q}^2}&=&\frac{1}{4\pi r}
\end{eqnarray*}
\begin{eqnarray*}
\int\frac{d^3k}{(2\pi)^3}\frac{1}{{\bf k}^2({\bf k-q})^2}&=&\frac{1}{8|{\bf q}|}\\
\int\frac{d^3k}{(2\pi)^3}\frac{k^i}{{\bf k}^2({\bf k-q})^2}&=&\frac{q^i}{16|{\bf q}|}\\
\int^\Lambda\frac{d^3k}{(2\pi)^3}\frac{k^ik^j}{{\bf k}^2({\bf k-q})^2}&\approx&\frac{|{\bf q}|}{64}\left(3\frac{q^iq^j}{{\bf q}^2}-\delta^{ij}\right)+\frac{\Lambda\delta^{ij}}{6\pi^2}.
\end{eqnarray*}

\chapter{Integrais em $4D$}
Aproximações $P^\mu=M\delta^\mu_0$, $|q^0|\ll |{\bf q}|\ll M$ e $E\ll M$, veja referências \cite{Bjerrum,Torma}
\begin{eqnarray*}
J&=&\int\frac{d^4k}{(2\pi)^4}\frac{1}{k^2(k-q)^2[(P-k)^2-M^2]}\approx\frac{-i}{32M|{\bf q}|}\\
J_\mu&=&\int\frac{d^4k}{(2\pi)^4}\frac{k_\mu}{k^2(k-q)^2[(P-k)^2-M^2]}\approx\frac{-iq_\mu}{64M|{\bf q}|}+\frac{i\delta^0_\mu}{16\pi^2M}\ln\frac{|{\bf q}|}{M}\\
J_{\mu\nu}&=&\int\frac{d^4k}{(2\pi)^4}\frac{k_\mu k_\nu}{k^2(k-q)^2[(P-k)^2-M^2]}\\&\approx&\frac{-i}{256M|{\bf q}|}[3q_\mu q_\nu+q^2(\delta^0_\mu\delta^0_\nu-\eta_{\mu\nu})]+\frac{i(q_\mu\delta^0_\nu+q_\nu\delta^0_\mu)}{32\pi^2M}\ln\frac{|{\bf q}|}{M}
\end{eqnarray*}
\begin{eqnarray*}
K+K'&=&\int\frac{d^4k}{(2\pi)^4}\frac{1}{(k^2-\mu^2)[(k-q)^2-\mu^2][(P-k)^2-M^2][(p+k)^2-m^2]}
\\&&+\int\frac{d^4k}{(2\pi)^4}\frac{1}{(k^2-\mu^2)[(k-q)^2-\mu^2][(P-k)^2-M^2][(p'-k)^2-m^2]}
\\&\approx&\frac{-1}{8\pi M{\bf q}^2|{\bf p}|}\ln\frac{|{\bf q}|}{\mu}
\end{eqnarray*}

\chapter{Deflexão clássica de partículas massivas e sem massa}
Neste apêndice vamos calcular os resultados clássicos para a deflexão gravitacional de partículas massivas e sem massa \cite{Collins,Fischbach,Accioly02} por um centro espalhador de massa $M$ descrito pela geometria de Schwarzschild
\[ds^2=\left(1-\frac{2GM}{r}\right)dt^2-\left(1-\frac{2GM}{r}\right)^{-1}dr^2-r^2d\theta^2-r^2\sin^2\theta d\phi^2.\]
As equações do movimento podem ser escritas como \cite{Wald}
\[\frac{\dot{r}^2}{2}+V(r)=\frac{E^2}{2}\;\;\;,\;\;\;L=r^2\dot{\phi}\;\;\;e\;\;\;\theta=\frac{\pi}{2}\]
onde
\[V(r)=\frac{1}{2}\left(1-\frac{2GM}{r}\right)\frac{L^2}{r^2}\]
para partículas sem massa e
\[V(r)=\frac{1}{2}\left(1-\frac{2GM}{r}\right)\left(\frac{L^2}{r^2}+1\right)\]
para partículas massivas.

Em ambos os casos, o ângulo de deflexão é dado por
\[\phi=2\int_{r_0}^{\infty}\frac{L/r^2}{\sqrt{E^2-2V(r)}}dr-\pi\]
onde $r_0$ é a menor distância alcançada pela partícula ao centro, dada pela maior raiz da equação $V(r_0)=E^2/2$, assim temos
\[r_0=\frac{2b}{\sqrt{3}}\cos{\left[\frac{1}{3}\cos^{-1}{\left(-\frac{3\sqrt{3}GM}{b}\right)}\right]}\]
e
\[r_0=\frac{2b}{\sqrt{3}}\sqrt{1+\frac{1}{3}\left(\frac{2GM}{b}\frac{1-v^2}{v^2}\right)^2}\times\]\[\cos{\left[\frac{1}{3}\cos^{-1}{\left(-\frac{3\sqrt{3}}{2}\frac{\frac{2GM}{b}+\frac{1}{3}\frac{2GM}{b}\frac{1-v^2}{v^2}+\frac{2}{27}\left(\frac{2GM}{b}\frac{1-v^2}{v^2}\right)^3}{\left(1+\frac{1}{3}\left(\frac{2GM}{b}\frac{1-v^2}{v^2}\right)^2\right)^{3/2}}\right)}\right]} \]
\[-\frac{2GM}{3}\frac{1-v^2}{v^2}\]
para partículas sem massa e massivas respectivamente e $b\equiv L/E$ é o parâmetro de impacto no caso sem massa e $E\equiv1/\sqrt{1-v^2}$ e $L\equiv bv/\sqrt{1-v^2}$ onde $v$ é a velocidade da partícula massiva no infinito.

Fazendo ainda a substituição na integral de $x=r_0/r$ e $\alpha=2GM/r_0$ obtemos
\[\phi=2\int_{0}^{1}\frac{dx}{\sqrt{(1-x^2)-\alpha(1-x^3)}}-\pi\]
e
\[\phi=2\int_{0}^{1}\frac{dx}{\sqrt{(1-x^2)-\alpha(1-x^3)-\frac{\alpha(1-\alpha)(1-v^2)}{1-(1-\alpha)(1-v^2)}(1-x)}}-\pi\]
expandindo em $GM/b\ll 1$ temos
\[\phi=4\frac{GM}{b}+\frac{15\pi}{4}\left(\frac{GM}{b}\right)^2+\frac{128}{3}\left(\frac{GM}{b}\right)^3+\frac{3465\pi}{64}\left(\frac{GM}{b}\right)^4+\ldots\]
e
\begin{eqnarray*}
\phi&=&\left(2+\frac{2}{v^2}\right)\frac{GM}{b}+\left(\frac{3\pi}{4}+\frac{3\pi}{v^2}\right)\left(\frac{GM}{b}\right)^2+\left(\frac{10}{3}+\frac{30}{v^2}+\frac{10}{v^4}-\frac{2}{3v^6}\right)\left(\frac{GM}{b}\right)^3\\&&+\left(\frac{105\pi}{64}+\frac{105\pi}{4v^2}+\frac{105\pi}{4v^4}\right)\left(\frac{GM}{b}\right)^4+\ldots
\end{eqnarray*}
é interessante notar que, apesar de seguirem geodésicas distintas, e portanto resultarem em diferentes ângulos de deflexão, o desvio para as partículas massivas no limite em que $v\rightarrow 1$ é idêntico ao caso das sem massa.

\chapter{Deflexão gravitacional pelo método de JWKB}
Neste apêndice vamos seguir de perto o trabalho de Sanchez \cite{Sanchez} sobre o espalhamento de ondas escalares na métrica de Schwarzschild.

Para isso vamos resolver pelo método de separação de variáveis a equação de onda para o escalar $\varphi(x)$
\[\Box\varphi(x)=g^{\mu\nu}\varphi(x)_{;\mu\nu}=0\]
onde a métrica de Schwarzschild em coordenadas padrão é dada por
\[ds^2=\left(1-\frac{2GM}{r}\right)dt^2-\left(1-\frac{2GM}{r}\right)^{-1}dr^2-r^2d\theta^2-r^2\sin^2\theta d\phi^2\]
e escrevendo $\varphi(x)$ como
\[\varphi(x)=\frac{R(r)}{r}Y_{lm}(\theta,\phi)e^{-i\omega t}\]
onde $Y_{lm}(\theta,\phi)$ é um harmônico esférico e ainda substituindo a coordenada $r$ pela coordenada (tortoise) de Regge-Wheeler
\[r^*=r+2GM\ln\left(\frac{r}{2GM}-1\right)\]
obtemos uma equação diferencial simples
\[\left[\frac{d^2}{dr^{*2}}+\omega^2-V_{ef}(r)\right]R(r^*)=0\]
onde
\[V_{ef}(r)=\left(1-\frac{2GM}{r}\right)\left(\frac{2GM}{r^3}+\frac{l(l+1)}{r^2}\right).\]

Esta equação possui solução assintótica $(r\rightarrow\infty)$ da forma \cite{Matzner}
\[R_l(r)\approx\sin[\omega r+2GM\omega\ln(2\omega r)+\delta_l(\omega)-l\pi/2]\]
onde $\delta_l(\omega)$ é o ``shift" da função de onda.

Por outro lado, a fase da função de onda pode ser calculada pelo método de JWKB \cite{Landau2} válido para pequenos comprimentos de onda (ou seja, $\omega b=\frac{2\pi b}{\lambda}\gg 1$)
\[\int_{r^*_0}^{r^*}\sqrt{\omega^2-V_{ef}(r^*)}dr^*+\pi/4\]
onde $r_0$ é o ponto de retorno clássico $\omega^2=V_{ef}(r_0)$, este método é válido na aproximação $(l+1/2)=\omega b$.

Voltando a coordenada $r$ usual através da relação
\[dr^*=\frac{dr}{1-\frac{2GM}{r}}\]
obtemos a seguinte expressão para o ``shift"
\[\delta_l(\omega)=\lim_{r\rightarrow\infty}\left[\int_{r_0}^r\frac{dr}{1-\frac{2GM}{r}}\sqrt{\omega^2-V_{ef}(r)}-\omega r+\frac{\pi}{2}\left(l+\frac{1}{2}\right)-2GM\omega\ln(2\omega r)\right].\]

A relação entre o ``shift" e a deflexão é dada por \cite{Landau2}
\begin{eqnarray*}
\theta&=&-2\frac{d\delta_l(\omega)}{dl}\\
&=&-\frac{\lambda}{\pi}\frac{d\delta_b(\lambda)}{db}\\
&=&-2\frac{d}{db}\int_{r_0}^\infty\frac{dr}{1-\frac{2GM}{r}}\sqrt{1-\left(1-\frac{2GM}{r}\right)\left[\frac{b^2}{r^2}+\frac{\lambda^2}{4\pi^2r^2}\left(\frac{2GM}{r}-\frac{1}{4}\right)\right]}-\pi\\&=&2b\int_{r_0}^\infty\frac{dr}{r^2\sqrt{1-\left(1-\frac{2GM}{r}\right)\left[\frac{b^2}{r^2}+\frac{\lambda^2}{4\pi^2r^2}\left(\frac{2GM}{r}-\frac{1}{4}\right)\right]}}-\pi\\&=&2\int_0^1\frac{dx}{\sqrt{(1-\alpha)[1+\beta(\alpha-1/4)]-x^2(1-\alpha x)[1+\beta(\alpha x-1/4)]}}-\pi
\end{eqnarray*}
onde substituímos $x=r_0/r$, $\alpha=2GM/r_0$, $\beta=\lambda^2/4\pi^2b^2$ e usamos a equação $\omega^2=V_{ef}(r_0)$.

Como estamos interessados na aproximação $GM/b\ll 1$ e $\lambda/b\ll 1$ podemos expandir em série em função destes dois parâmetros e assim obtemos
\[r_0=b\left(1-\frac{GM}{b}-\frac{\lambda^2}{32\pi^2b^2}-\frac{3G^2M^2}{2b^2}+\frac{GM\lambda^2}{4\pi^2b^3}-\frac{\lambda^4}{2048\pi^4b^4}+\ldots\right)\]
e portanto \cite{Accioly6}
\begin{eqnarray*}
\theta&=&\frac{4GM}{b}+\frac{\lambda^2}{32\pi b^2}+\frac{15\pi G^2M^2}{4b^2}-\frac{3GM\lambda^2}{4\pi^2b^3}+\frac{3\lambda^4}{2048\pi^3b^4}\\
&&+\frac{128G^3M^3}{3b^3}-\frac{99G^2M^2\lambda^2}{128\pi b^4}-\frac{7GM\lambda^4}{64\pi^4 b^5}+\frac{5\lambda^6}{65536\pi^5 b^6}+\ldots
\end{eqnarray*}
observe que esta expressão retorna ao resultado clássico do apêndice anterior no limite geométrico $(\lambda\rightarrow 0)$.

\end{document}